\documentclass[aps,twocolumn,floatfix]{revtex4-1}

\usepackage{amssymb}
\usepackage{graphicx}
\usepackage{caption}
\usepackage{multirow}
\usepackage[table,xcdraw]{xcolor}
\usepackage{soul}
\usepackage{kantlipsum}
\usepackage{amsmath}
\usepackage{amsfonts}
\usepackage{amssymb}
\usepackage{dcolumn} 
\usepackage{braket}
\usepackage{booktabs}
\usepackage{multirow}
\usepackage[table]{xcolor}
\usepackage{dsfont}
\usepackage{times}
\usepackage{epsfig}
\usepackage{xcolor}
\usepackage{tikz}
\usepackage{float}
\usepackage{color}
\usepackage{inputenc}
\usepackage{booktabs}
\usepackage{pgfplotstable}
\def\bx {{\bf x}}
\def\by {{\bf y}}

\def\bkappa {\boldsymbol{\kappa}}

\newcommand{\beq}{\begin{equation}}
\newcommand{\eeq}{\end{equation}}
\newcommand{\ba}{\begin{eqnarray}}
\newcommand{\ea}{\end{eqnarray}}
\newcommand{\matA}{\underline{\mathbf{A}}}
\newcommand{\matB}{\underline{\mathbf{B}}}

\input epsf

\usepackage[english]{babel}

\begin{document}
\title[]{\hspace{2cm} Tunable three-way valley Hall energy-splitter: \newline venturing beyond graphene-like structures}

\author{Mehul P. Makwana$^{1, 2}$ and Gregory J. Chaplain$^{1}$}
\affiliation{$^1$ Department of Mathematics, Imperial College London, London SW7 2AZ, UK }
\affiliation{$^2$ Multiwave Technologies AG, 3 Chemin du Pr\^{e} Fleuri, 1228, Geneva, Switzerland}

\begin{abstract}
Strategically combining four structured domains creates the first ever three-way topological energy-splitter; remarkably, this is only possible using a square, or rectangular, lattice, and not the graphene-like structures more commonly used in valleytronics. To achieve this effect, the two mirror symmetries, present within all fully-symmetric square structures, are broken; this leads to two nondistinct interfaces upon which valley-Hall states reside. These interfaces are related to each other via the time-reversal operator and it is this subtlety that allows us to ignite the third outgoing lead. The geometrical construction of our structured medium allows for the three-way splitter to be adiabatically converted into a wave steerer around sharp bends. Due to the tunability of the energies directionality by geometry, our results have far-reaching implications for applications such as beam-splitters, switches and filters across wave physics.
\end{abstract}
\maketitle


\subsection*{Introduction}
\label{sec:intro}

A fundamental understanding of the manipulation and channeling of wave energy underpins advances in device design in acoustics and optics \cite{mekis_high_1996, yariv_coupled-resonator_1999, chutinan_wider_2002}. For instance, beam-splitters, that split an incident beam of light in two, are extensively used for experiments and devices in quantum computing, astrophysics, relativity theory and other areas of physics \cite{quirrenbach_optical_2001, kok_linear_2007, mitomi_design_1995}. This desire to guide waves, split and redirect them, for broadband frequencies, in a lossless and robust manner, extends well beyond optical devices and into electromagnetism, vibration control and acoustic switches, amongst other fields \cite{ju_topological_2015, liu_multimode_2004, ma_guiding_2015}. Fortunately, the advent of topological insulators in quantum mechanics  \cite{kane_z2_2005, xiao_valley-contrasting_2007}, and their translation into classical systems, has led to waveguides that are more broadband and robust than previous designs \cite{gao_valley_2017, lu_observation_2016, shalaev_experimental_2017, ma_all-si_2016, makwana_geometrically_2018} and ultimately to  robust networks \cite{cheng_robust_2016, wu_direct_2017, xia_topological_2017, zhang_manipulation_2018, qiao_electronic_2011};
however, the vast majority of the topological energy-splitters  are based upon graphene-like hexagonal structures and hence restricted to a two-way partitioning of energy. Herein we rectify this with an intelligently engineered three-way topological energy-splitter, the geometrical design of which is based upon the square lattice \cite{he_emergence_2015, xia_observation_2018}. 

\begin{figure}[!ht]
    \centering
    \begin{minipage}{.25\textwidth}
                \hspace{-1.6cm}
        \includegraphics[width=0.69\linewidth]{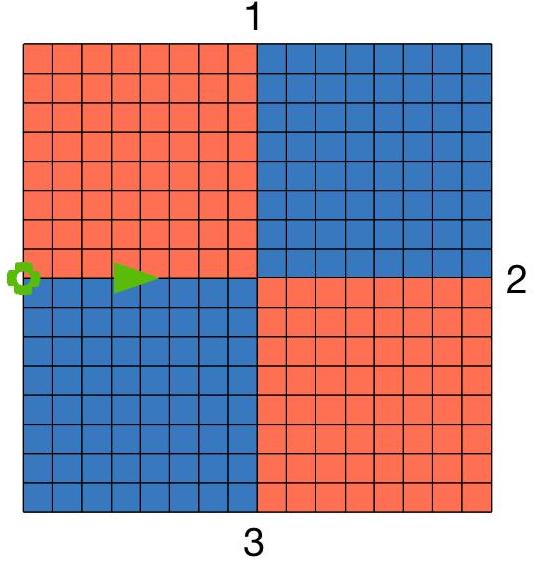}
        \label{fig:prob1_6_2}
    \end{minipage}%
    \begin{minipage}{0.10\textwidth}
         \hspace{-3.65cm}
        \includegraphics[width=0.70\linewidth]{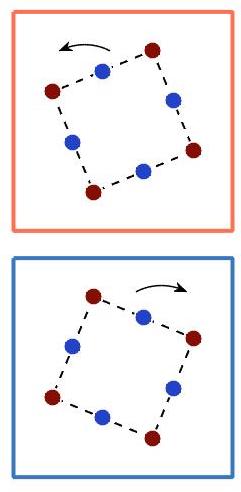} 
        \label{fig:prob1_6_1}
    \end{minipage}
        \begin{minipage}{0.10\textwidth}
         \hspace{-2.10cm}
        \includegraphics[width= 1.70\linewidth]{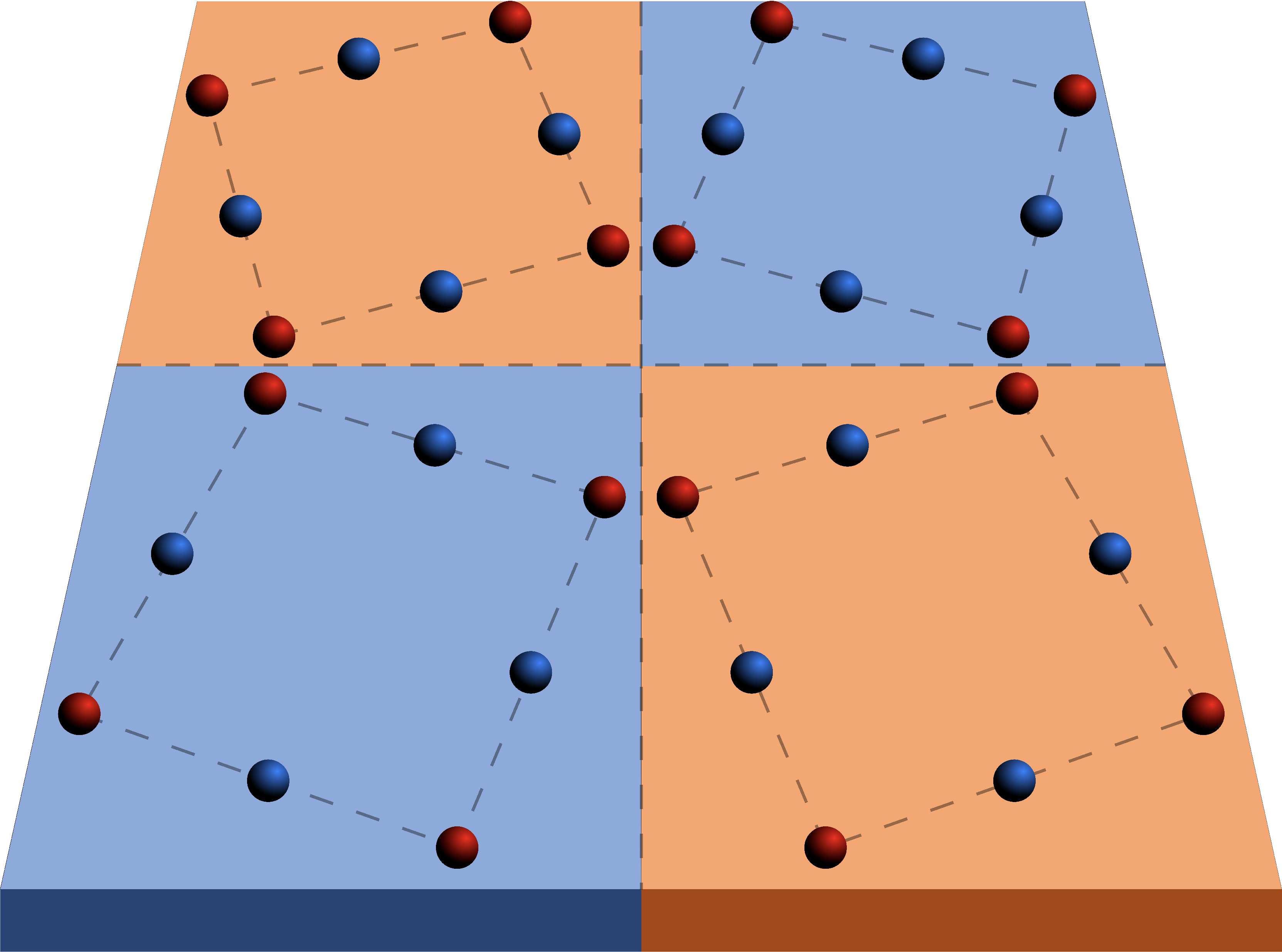} 
        \label{fig:prob1_6_1}
    \end{minipage}   
        \begin{minipage}{.5\textwidth}
                 \hspace{-0.85 cm}
        \includegraphics[width=0.75\linewidth]{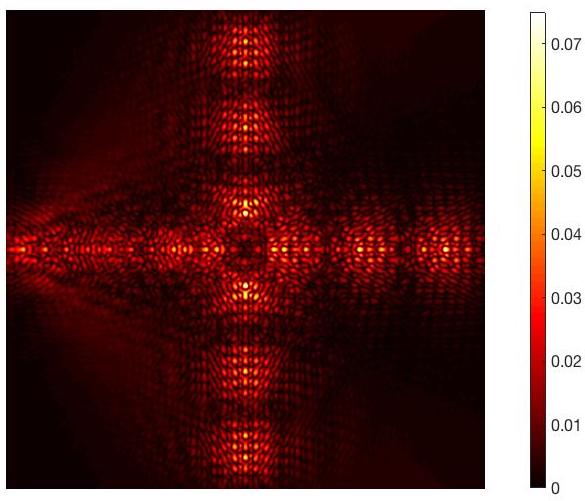} 
        \label{fig:prob1_6_2}
    \end{minipage}%
     \hspace{-0.70cm}
            \begin{minipage}{.15\textwidth}
        \includegraphics[width=0.80\linewidth]{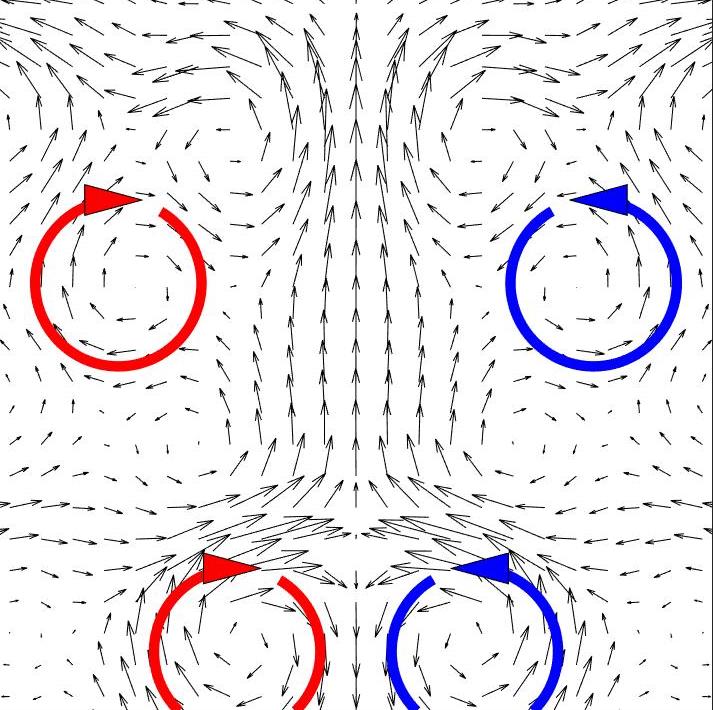}
        \label{fig:prob1_6_2}
        \caption*{Exit 1}
    \end{minipage}%
            \begin{minipage}{.15\textwidth}
        \includegraphics[width=0.80\linewidth]{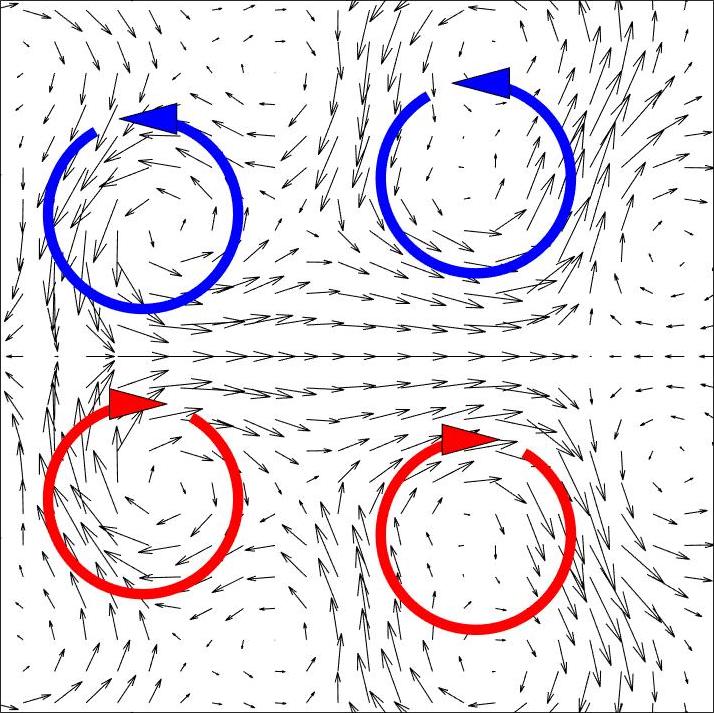}
        \label{fig:prob1_6_2}
        \caption*{Exit 2}
    \end{minipage}%
            \begin{minipage}{.15\textwidth}
        \includegraphics[width=0.80\linewidth]{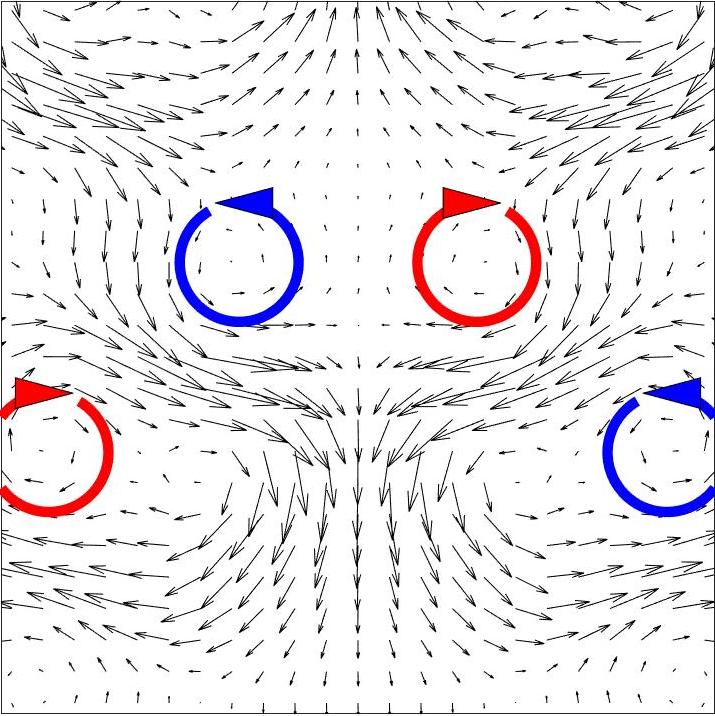}
        \label{fig:prob1_6_2}
        \caption*{Exit 3}
    \end{minipage}%
    \caption{Three-way topological splitter ($\omega = 6.5356$) \textemdash \emph{Top-left}: the canonical splitter geometry of four structured quadrants. Different orientation of scatterers in the orange and blue regions. Source is indicated by circle at left edge. \emph{Top-right}: zoom-in of nodal region cells.  \emph{Middle}: displacement field, illustrating the splitting of energy three ways. \emph{Lower}: opposite chirality at the interfaces.}
    \label{fig:3_way_splitter}
\end{figure}

Time-reversal symmetric (TRS) topological guides leverage the discrete valley degrees of freedom that arise from degenerate extrema in Fourier space. When constructing topological guides, graphene-like materials are the prime candidate due to their well-defined $KK'$ valleys; these valleys are distinguished by their opposite 
chirality and related by TRS.  The intervalley scattering is heavily suppressed \cite{chen_defect_2009, morozov_strong_2006, morpurgo_intervalley_2006, makwana_designing_2018} by the  large Fourier separation between the two valleys, and each valley becomes an efficient information carrier. These valley modes are attracting growing attention, in part due to their simplicity of construction, leading to the emergent field of valleytronics  \cite{xiao_valley-contrasting_2007, gao_valley_2017, lu_observation_2016, shalaev_experimental_2017, ma_all-si_2016, makwana_geometrically_2018}. The primary benefits of these topologically nontrivial modes over,  cavity and topologically trivial interfacial modes \cite{makwana_designing_2018}, is the additional topological protection afforded by the chiral flux either side of the zero-line modes (ZLMs) and geometrical tunability \cite{makwana_designing_2018} allowing a bend to be adiabatically converted into a splitter (and vice-versa). 

The prevalence of graphene-like structures has primarily limited valleytronic devices to \emph{two-way} energy-splitters; this is motivated  by the conservation of 
 chirality at the $KK'$ valleys \cite{cha_experimental_2018, he_acoustic_2016, he_silicon--insulator_2018, he_two-dimensional_2019, khanikaev_two-dimensional_2017, nanthakumar_inverse_2019, ozawa_topological_2019, qiao_current_2014, schomerus_helical_2010, shen_valley-projected_2019, xia_topologically_2019, yan_-chip_2018, ye_observation_2017, cheng_robust_2016, wu_direct_2017, xia_topological_2017, zhang_manipulation_2018, qiao_electronic_2011}. A four-way partitioning of energy away from a nodal \emph{region} was shown in \cite{makwana_designing_2018} however this was dependent upon the tunneling mechanism. Tunneling would introduce an additional dependency upon the system; namely, the decay length perpendicular to the direction of propagation. Hence, the transmission along the outgoing leads would be heavily contingent upon the location of the mode within the topologically nontrivial band-gap; therefore an alternative method whereby the energy is partitioned away from a well-defined nodal \emph{point} as opposed to a nodal \emph{region} is highly desirable. Importantly, this is only possible using a square or rectangular lattice; the three-way energy splitting is dependent upon the equivalence of the interfaces (modulo time-reversal symmetry) that is only achievable using the four-fold symmetric cellular structure. The \emph{geometrical tunability}, the \emph{topological robustness} and the \emph{three-way} partitioning of energy away from a well-defined nodal \emph{point} are three crucial advantages of the square energy-splitter over competing designs. 

We begin in Sec. \ref{sec:system} by explicitly recasting the continuum plate model into the language of quantum mechanics, utilising a Hamiltonian description, while retaining elements of the continuum language to bridge across the quantum and elastic plate communities. Despite us utilising the structured elastic plate equation, our theories are system independent, hence are transposable to other classical systems. We examine a square cellular structure containing only a single mirror symmetry in Sec. \ref{sec:C2v}; we demonstrate how this restricts a medium, comprised of these cells, to solely yield straight valley-Hall guides i.e. the energy cannot be navigated around a bend. Contrastingly, the structure examined in Sec. \ref{sec:C4v} contains two mirror symmetries which in turn allows for ZLMs to couple around a bend as well as partition three-ways away from a nodal point. A few concluding remarks are drawn together in Sec. \ref{sec:conclusion}.

\section{Formulation}
\label{sec:system}
\subsection{Kirchhoff-Love plate}
The group theoretic and topological concepts foundational to our approach hold irrespective of any specific two-dimensional scalar wave system. We choose to illustrate them here using a structured thin elastic Kirchhoff- Love (K-L) plate \cite{landau_theory_1970} for which many results for point scatterers are explicitly available \cite{evans_penetration_2007}; the geometrical ideas themselves carry across to photonics, phononics and plasmonics. Displacement Bloch eigenstates  $ \ket{\psi_{n \bkappa}}$ satisfy the 
(non-dimensionalised) K-L equation,
\beq
\hat{H} \ket{\psi_{n \bkappa}} = \omega_{n\bkappa}^2
\ket{\psi_{n \bkappa}} + F(\bx) \ket{\psi_{n \bkappa}}, 
\label{eq:simult_equations}\eeq  for Bloch-wavevector ${\bkappa}$, $n$ denoting the eigenmodes and
 $\omega_{n, \bkappa}$ the non-dimensionalised frequency; 
 reaction forces at the point constraints, $F(\bx)$, introduce
 dependence upon the direct lattice. The most straightforward constraints, sufficient for our purposes, are point mass-loading with 
the reaction forces proportional to the displacement 
 via an effective impedance coefficient such that
\beq
F(\bx)=\omega_{n \bkappa}^2\sum_{\bf l}\sum_{p=1}^{P}
 M^{(p)}_{\bf l} \psi_{n \bkappa}(\bx)\delta\left({\bf x}-{\bf x}^{(p)}_{\bf l}\right).
\label{eq:ML_RHS}
\eeq Here ${\bf l}$ labels each elementary cell that repeats 
periodically to create the infinite physical plate crystal, and
each cell contains $p=1...P$ constraints. The component equation is retrieved from
Eq. (\ref{eq:simult_equations}) using 
\beq
\hat{H} = \int \ket{\bx} H(\bx, \by) \bra{\by} d\bx d\by, \quad H(\bx, \by) = \delta(\bx - \by)\nabla^4_{\bx}.
\nonumber
\eeq In an infinite medium the displacements are Bloch eigenfunctions 
\beq
\psi_{n \bkappa}(\bx) = \braket{\bx|\psi_{n\bkappa}} = \exp\left(i\bkappa\cdot\bx \right) \braket{\bx|u_{n\bkappa}},
\label{eq:Bloch_condition}
\eeq where $\ket{u_{n\bkappa}}$ is a periodic eigenstate.
The displacements satisfy the following completeness and orthogonality relations: 
\beq
\sum_{n \bkappa} \ket{\psi_{n\bkappa}} \bra{\psi_{n\bkappa}} = \hat{1}, \quad \braket{\psi_{n\bkappa}|\psi_{m\bkappa'}} = \delta_{mn} \delta_{\bkappa, \bkappa'}.
\label{eq:complete_orthogonal_set}
\eeq  Applying the Bloch conditions to the relations \eqref{eq:complete_orthogonal_set} yields an identical completeness relation and the following orthogonality condition, 
\beq
\int u^*_{n\bkappa}u_{m\bkappa'} \exp\left(i \Delta\bkappa' \right) d\bx= \delta_{mn} \delta_{\bkappa, \bkappa'}.
\label{eq:u_orthogonal}
\eeq where $\Delta\bkappa' = \bkappa'-\bkappa$. Due to the completeness of the periodic eigensolution, we can expand $\ket{u_{n\bkappa}}$ in the complete orthogonal basis set $\{u_{j\bkappa_0}(\bx) \}$ where $\bkappa_0$ is fixed,
\begin{multline}
\ket{\psi_{n\bkappa}} =  \exp\left(i\bkappa\cdot\bx \right) \ket{u_{n\bkappa}} = \\
=  \exp\left(i\bkappa\cdot\bx \right) \sum_m A_{nm}(\bkappa) \ket{u_{m\bkappa_0}}  = \\
\exp\left(i\Delta\bkappa\cdot\bx \right) \sum_m A_{nm}(\bkappa) \ket{\psi_{m\bkappa_0}},
\label{eq:psi_expansion}
\end{multline} where $\Delta\bkappa = \bkappa - \bkappa_0$.  After substituting \eqref{eq:psi_expansion}, into the governing equation \eqref{eq:simult_equations}, we explicitly obtain,
\begin{multline}
\exp\left(i\Delta\bkappa\cdot\bx \right)\sum_m A_{nm}(\bkappa) [(\omega_{m\bkappa_0}^2-\omega_{n\bkappa}^2) \times \\
\left[1 + \sum_{{\bf N}, p}
 M^{(p)}_{\bf N} \delta\left({\bf x}-{\bf x}^{(p)}_{\bf N}\right) \right]
+4i \Delta\bkappa \cdot \nabla^3_{\bx}  \\
+ \mathcal{O} \left(|\Delta\bkappa|^2 \right)]\psi_{j\bkappa_0}(\bx)=0.
\label{eq:expansion_gov}\end{multline} up to first-order in $|\Delta\bkappa|$. This expansion will be used in the subsequent section, alongside symmetry considerations,  to engineer the Dirac cones.

\section{$C_{2v}$ cellular structure}
\label{sec:C2v}

\begin{figure}
\begin{tabular}{cc}
	 (a) & (b) \\
		\includegraphics[height=0.45\linewidth]{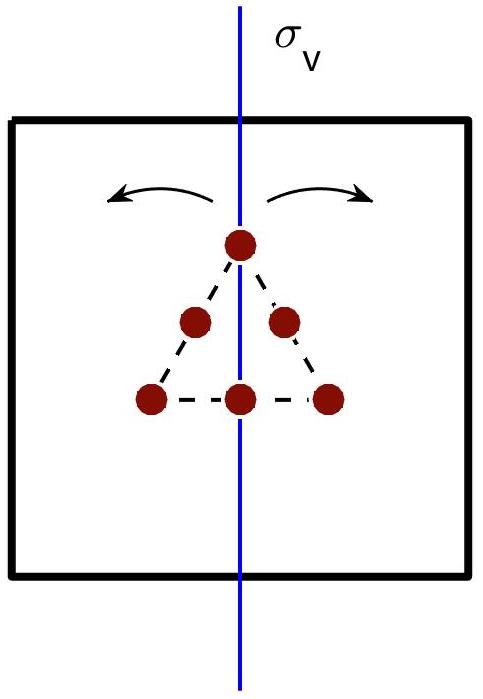} &
		\includegraphics[height=0.395\linewidth]{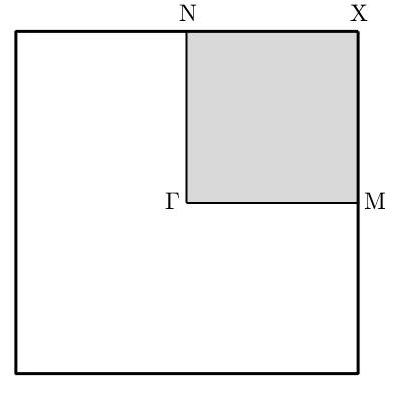}
\end{tabular}
\caption{(a) cellular structure shown; uniform mass values of $1$, lattice constant of $2$, centroid to vertex mass distance  of $0.45$. Pre-perturbation structure has $\sigma_v$ symmetry, post-perturbation structure breaks  $\sigma_v$ symmetry via an angular perturbation of the inclusion set. Panel (b) shows the irreducible Brillouin zone (IBZ, shaded region) within the Brillouin zone (BZ).}
\label{fig:C2v_case}
\end{figure}

In this section we examine the cellular structure shown in Fig. \ref{fig:C2v_case}(a). Spatially, this structure solely has $\sigma_v$ reflectional symmetry; however, in Fourier space, it has $C_{2v}$ symmetry, due to the presence of time-reversal symmetry. In subsection \ref{sec:Dirac_formation}, we utilise the expansion \eqref{eq:expansion_gov} and group theoretical considerations to demonstrate how an accidental Dirac cone is engineered. The effects of $\sigma_v$ symmetry breaking, on the bulk bandstructure, are discussed in 
subsection \ref{sec:C2v_bulk}. Subsection \ref{sec:C2v_ribbons} demonstrates how the strategic stacking of geometrically distinct media results in valley-Hall edge states \cite{makwana_geometrically_2018, xiao_valley-contrasting_2007}. A ZLM connected to a valley-Hall edge state is shown in subsection \ref{sec:C2v_scattering} alongside a justification for why this particular $C_{2v}$ model does not allow for propagation around corners. Since the valley-Hall state is a weak topological state protected solely by symmetry, care must be taken to prohibit backscattering hence knowledge of the long-scale envelope is especially useful for finite length interfaces as it can used to minimise the backscattering. An asymptotic method, more commonly known as high-frequency homogenisation (HFH) allows for the characterisation of this long-scale envelope (Appendix); this is  applied to a $C_{2v}$ ZLM in subsection \ref{sec:C2v_scattering}. 

\subsection{Engineering an accidental Dirac cone}
\label{sec:Dirac_formation}
\subsubsection*{Band coupling at high-symmetry point for $C_{2v}$ structure}
The point group symmetry of the structure, shown in Fig. \ref{fig:C2v_case}(a), is $G_{\Gamma} = C_{2v}$; this is also the point group symmetry at $N$, $G_N = C_{2v}$ (Table \ref{table:c2v_table}). The $C_{2v}$ point group arises from a combination of spatial (reflectional) and time-reversal symmetries; the latter relates $\bkappa \rightarrow -\bkappa$. The group theoretical arguments used throughout this subsection, are reminiscent of those found in \cite{he_emergence_2015} although in our calculations we have applied an actual asymptotic scheme whereby we have judiciously chosen a small parameter with a distinguished limit. 

The irreducible representations (IRs) at  $N$ are one-dimensional hence there is no symmetry induced degeneracy. Despite this, we shall demonstrate in this subsection how two of the IRs can be tuned such that an accidental degeneracy (that is not symmetry repelled) forms. 
\\

\begin{figure} [h!]
	\centering
		\includegraphics[width=0.45\textwidth]{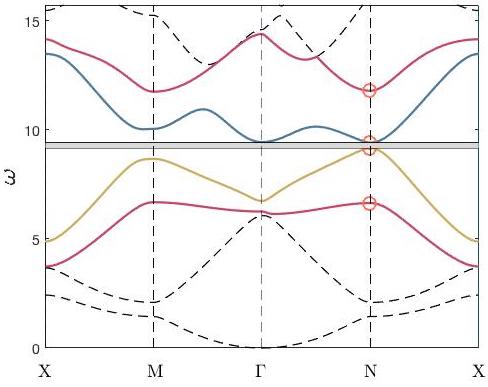}
	\caption{Dispersion curves $C_{2v}$ case (when $\omega_{A_1}>\omega_{B_1}$ at $N$) \textemdash Parameter values are different to those in Fig. 2; $\sigma_v$ symmetry  present within physical space cell. Parameter values: distance between centroid and vertex mass $=0.45$, lattice constant $=2$, vertex mass value $=1$, non-vertex mass value $=0.5$. The coloured bands are associated with the SSE. In this instance, $A_1$ curve lies above $B_1$ curve at $N$, hence there is no band crossing along $NX$.} 
\label{fig:dc_sse}
\end{figure}

\begin{figure}
    \begin{minipage}{.25\textwidth}
    \caption*{IR: $B_2$, Basis: $y$}
        \includegraphics[width=0.70\linewidth]{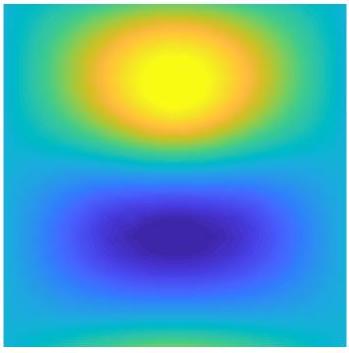}
        \label{fig:prob1_6_2}
    \end{minipage}%
    \begin{minipage}{0.25\textwidth}
        \hspace{-1.85cm}
               \caption*{IR: $B_1$, Basis: $x$}
        \includegraphics[width=0.70\linewidth]{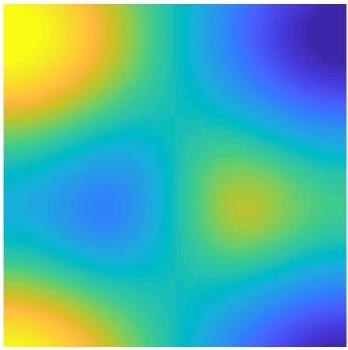}
        \label{fig:prob1_6_1}
    \end{minipage}
    \begin{minipage}{0.25\textwidth}
        \caption*{IR: $A_1$, Basis: $x^2, y^2 $}
        \includegraphics[width=0.70\linewidth]{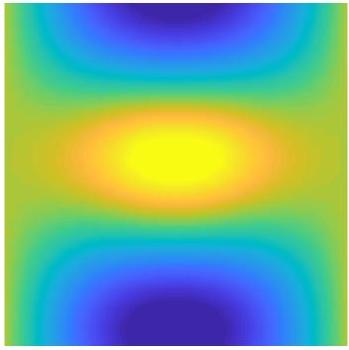}
        \label{fig:prob1_6_2}
    \end{minipage}%
    \begin{minipage}{0.25\textwidth}
        \hspace{-1.85cm}
        \caption*{IR: $A_2$, Basis: $xy$}
        \includegraphics[width=0.70\linewidth]{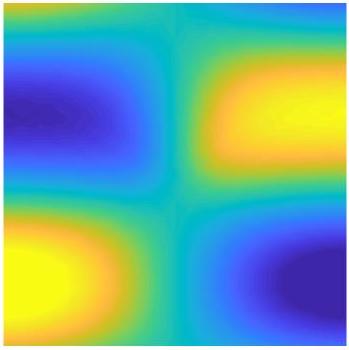}
        \label{fig:prob1_6_1}
    \end{minipage}
    \caption{Eigensolutions, at $N$, for $C_{2v}$ case with $\sigma_v$ symmetry.}
        \label{fig:prepert_orbs}
\end{figure}

The four solid bands in Fig \ref{fig:dc_sse} (bands numbered $3-6$ inclusive) are associated with the eigensolutions, shown in Fig. \ref{fig:prepert_orbs}, these match the basis function symmetries of the the $C_{2v}$ group (Table \ref{table:c2v_table}); hence this indicates that bands $3-6$ are symmetry induced and the sequential ordering of them (lowest to highest) is deduced numerically, via the eigensolutions, as: $\{B_2, A_1, B_1, A_2\}$. 

\begin{table}[h!]
\begin{tabular}{|l|cccc|c|}
\hline
\cellcolor[HTML]{EFEFEF}Classes $\rightarrow$ &                       &                         &                              &                              &                                                                              \\
\cellcolor[HTML]{EFEFEF}IR $\downarrow$       & \multirow{-2}{*}{$E$} & \multirow{-2}{*}{$C_2$} & \multirow{-2}{*}{$\sigma_v$} & \multirow{-2}{*}{$\sigma_h$} & \multirow{-2}{*}{\begin{tabular}[c]{@{}c@{}}Basis \\ functions\end{tabular}} \\ \hline
$A_1$                                         & $+1$                  & $+1$                    & $+1$                         & $+1$                         & $x^2, y^2$                                                                   \\
$A_2$                                         & $+1$                  & $+1$                    & $-1$                         & $-1$                         & $xy$                                                                         \\
$B_1$                                         & $+1$                  & $-1$                    & $+1$                         & $-1$                         & $x, xy^2$                                                                    \\
$B_2$                                         & $+1$                  & $-1$                    & $-1$                         & $+1$                         & $y, x^2y$                                                                    \\ \hline
\end{tabular}
\caption{$C_{2v}$ character table}
\label{table:c2v_table}
\end{table}

It is expected, from the dispersion curves (Fig \ref{fig:dc_sse}) that the two bands that form the accidental degeneracy, namely $A_1, B_1$, have a strong influence on each other whilst the other two symmetry induced bands, $B_2, A_2$, will have a limited effect on the local curvature or slope of the $A_1, B_1$ bands \cite{dresselhaus_group_2008}; the effect, by the bands that lie outside of bands $3-6$, on  the $A_1, B_1$ bands, is expected to be negligible; to see these points mathematically we initially separate out the eigenket expansion Eq. \eqref{eq:psi_expansion} into two sets of bands; namely, the symmetry set eigensolutions (SSE), bands $3-6$, and those that lie outside the SSE,
\begin{multline}
\ket{\psi_{n\bkappa}} =  
\exp\left(i\Delta\bkappa\cdot\bx \right) \bigg[ \sum_{j \in \text{SSE}} A_{nj}(\bkappa) \ket{\psi_{j\bkappa_0}} + \\
\sum_{\alpha \notin \text{SSE}} A_{n\alpha}(\bkappa) \ket{\psi_{\alpha\bkappa_0}}  \bigg]. 
\label{eq:psi_separation} 
\end{multline} Motivated by the orthogonality condition \eqref{eq:u_orthogonal} and the expansion \eqref{eq:psi_separation}, we multiply equation \eqref{eq:psi_expansion}, by $\psi_{l \bkappa_0}^* (\bx)$ or $\psi_{\beta \bkappa_0}^* (\bx)$ (where $l \in \text{SSE}, \beta \notin \text{SSE}$) before integrating over the primitive cell to obtain the following two equations,
\begin{multline}
(\omega_{n\bkappa}^2-\omega_{l\bkappa_0}^2)\Lambda_{l}A_{nl} =  \sum_{j}H_{lj}A_{nj} + \sum_{\alpha}H_{l\alpha}A_{n\alpha},  \\
(\omega_{n\bkappa}^2-\omega_{\beta\bkappa_0}^2)A_{n\beta} =  \sum_{j}H'_{\beta j}A_{nj} + \sum_{\alpha}H'_{\beta \alpha}A_{n\alpha}
\label{eq:simul_eqns}
\end{multline} where the $\bkappa$ dependence of the weighting coefficients has been dropped; $\Lambda_{l}, H_{ab}, H'_{ab}$ are explicitly,
\begin{multline}
 \Lambda_{l}=1 + \sum_{p} M^{(p)}_{{\bf I}} \bigg|\psi_{l\bkappa_0}\left(\bx^{(p)}_{{\bf I}}\right)\bigg|^2, \\
H_{ab} = \Delta\bkappa \cdot \braket{\psi_{a\bkappa_0}|\hat{\bf{p}}_{1}|\psi_{b\bkappa_0}} + \mathcal{O}\left(|\Delta\bkappa|^2 \right), \\
H'_{ab} = H_{ab}+ \delta_{ab}(\omega_{n\bkappa}^2-\omega_{b\bkappa_0}^2) \times \\
 \sum_{p} M^{(p)}_{{\bf I}} \psi^*_{a\bkappa_0}\left(\bx^{(p)}_{{\bf I}}\right) \psi_{b\bkappa_0}\left(\bx^{(p)}_{{\bf I}}\right) + \mathcal{O}\left(|\Delta\bkappa|^2 \right). 
\label{eq:aug_Ham} 
\end{multline} Rearranging the second equation in \eqref{eq:simul_eqns} to, 
\beq
A_{n\beta} = \frac{\sum_{j}H_{\beta j}A_{nj} + \sum_{\alpha}H'_{\beta \alpha}A_{n\alpha}}{(\omega_{n\bkappa}^2-\omega_{\beta\bkappa_0}^2)}
\label{eq:Abeta_sub}\eeq and substituting this into the second summation of the first equation gives,
\begin{multline}
(\omega_{n\bkappa}^2-\omega_{l\bkappa_0}^2)\Lambda_{l}A_{nl} =  \sum_{j}H_{lj}A_{nj} + \\
\sum_{j} A_{nj} \sum_{\alpha} \frac{H_{l\alpha} H_{\alpha j}}{(\omega_{n\bkappa}^2-\omega_{\alpha\bkappa_0}^2)},
\label{eq:total_equation}\end{multline}
where we have neglected terms which couple states outside the SSE to other states outside the SSE. If we let $n=l \in \text{SSE}$ and $\bkappa = \bkappa_0 + \Delta\bkappa$ then the frequency term on the left-hand side is expanded to yield,
\beq
\omega_{n\bkappa}^2 = \omega_{l\bkappa_0}^2 + 2\omega_{l\bkappa_0} \Delta\bkappa \cdot \nabla_{\bkappa} \omega_{l \bkappa_0} + \mathcal{O}\left(|\Delta\bkappa|^2\right).
\label{eq:freq_exp}\eeq Hence, from this expansion it is easy to see that the second summation in  \eqref{eq:total_equation}, that couples states within the SSE to those outside, falls into second-order hence the effective first-order equation is, 
\beq
 \left(2\omega_{l\bkappa_0} \Delta \omega_l \Lambda_{l} \right)A_{nl} = \sum_{j \in \text{SSE}}H_{lj}A_{nj}, 
\label{eq:effective_eqn}\eeq where $ \Delta \omega_l =\omega_{l\bkappa} - \omega_{l\bkappa_0}$ and $l \in \text{SSE}$. Notably, the higher-order corrections, that encompass the coupling between bands within the SSE to those outside, provide the band curvature details away from a locally linear point. In this instance, Eq. \eqref{eq:effective_eqn} is a $4\times4$ matrix eigenvalue problem, where the Hamiltonian, with components $H_{lj}$, is Hermitian. If, for a particular $\bkappa_0$, the first-order term is zero we would have to proceed to second-order; here additional terms would come from the fourth-ordered derivative, the $\omega_{l\bkappa}$ expansion and band coupling between outside SSE and inside SSE bands.

\subsubsection*{Compatibility relations and band tunability along $NX$}
Bands tend to vary continuously except possibly at accidental degeneracies where modal inversion may occur which in turn leads to a discontinuity of the intersecting surfaces. Hence, the eigenfunctions continuously transform as you progress along a continuous IBZ path of simple eigenvalues. The associated IRs, that describe the transformation properties of the eigenfunctions, themselves smoothly transition into IRs that belong to the point groups along $N\Gamma$ or $NX$.  
\\

\begin{figure}
	\begin{tabular}{cc}
		(a) & (b) \\
		
		\includegraphics[height = 0.5\linewidth]{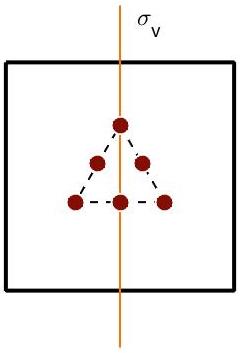} &
		\includegraphics[height = 0.5\linewidth]{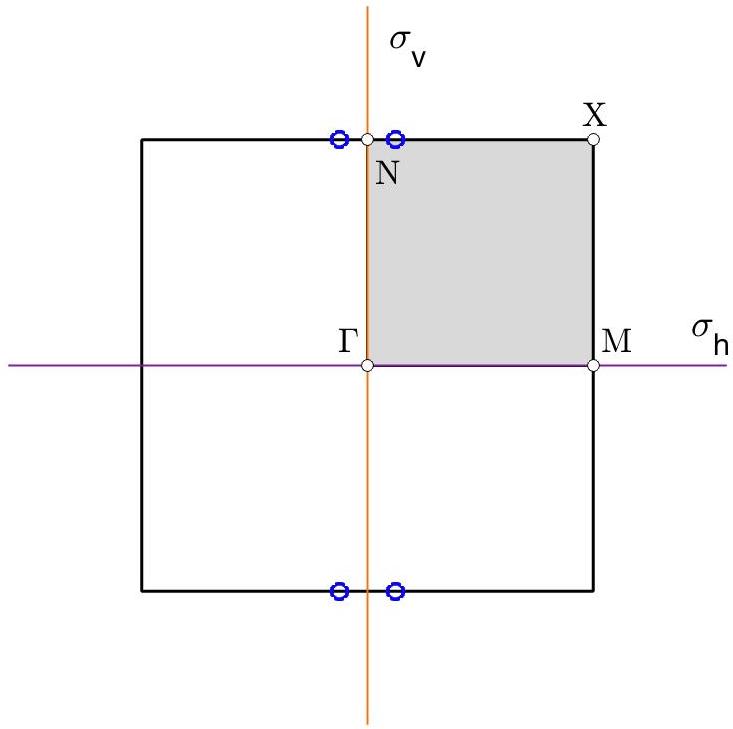} \\
		\multicolumn{2}{c}{(c)} \\
		\multicolumn{2}{c}{ \includegraphics[width=0.75\linewidth]{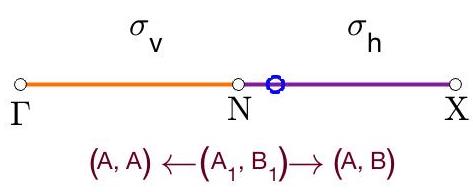}}
		
	\end{tabular}
	\caption{Physical and Fourier space cells \textemdash (a) cellular structure in physical space. (b) IBZ (shaded region) shown within BZ. Presence of $\sigma_v$ symmetry in physical space translates into $\sigma_h$ symmetry in Fourier space, this explains the symmetrical placement of the Dirac cones (blue circles) either side of $\sigma_h$.  (c) Unfurled IBZ path. Symmetries and IRs, along the paths $\Gamma \rightarrow N \rightarrow X$, are shown.}
	\label{fig:IBZ_path}
\end{figure}

 In physical space the cellular structure only has $\sigma_v$ spatial symmetry, this is equivalent to $\sigma_h$ symmetry in Fourier space, Fig. \ref{fig:IBZ_path}(b). Recall the definition of a point group symmetry, i.e. any symmetry operator $\hat{R} \in G_{\Gamma}$ that satisfies, $\hat{R} \bkappa = \bkappa \mod {\bf G}$, where ${\bf G}$ is a reciprocal lattice basis vector; this implies that $\bkappa \in NX$ solely has the mirror symmetry operator, $\sigma_h$ within its point group. Similarly, for a $\bkappa \in \Gamma N$, only the vertical mirror symmetry operator, $\sigma_v \in C_{2v}$, satisfies the point group criterion. The symmetries of the eigenfunctions, for a $\bkappa$ belonging to either of the paths, $NX$ and $N\Gamma$, are shown within the basis functions column of Table \ref{table:c2_table}. If we solely consider the two strongly coupled bands, represented by the IRs $A_1$ and $B_1$, then the associated eigenfunctions transitional behaviour, away from $N$, is described by the $\sigma_{v, h}$ character table. Due to the continuity of the bands the $A_1, B_1$ IRs belonging to the $C_{2v}$ table will transform into the IRs, of the $\sigma_{v,h}$ table, as we move away from $N$; the relationships between different IRs are more commonly referred to as compatibility relations \cite{dresselhaus_group_2008, heine_group_nodate}. Initially, we consider symmetry $\sigma_h$, the eigenstates at $N$ and along $NX$ satisfy the following,
\beq
\hat{P}_{\sigma_h} \ket{\psi_{A_1, B_1}} = \pm  \ket{\psi_{A_1, B_1}}, \quad \hat{P}_{\sigma_h} \ket{\psi_{A, B}} = \pm  \ket{\psi_{A, B}}.
\label{eq:sigma_h} \eeq  Hence, the bands $(A_1, B_1)$ (at $N$) are compatible with $(A, B)$ (along $NX$). Physically, this transition is also evident from the eigensolutions; as $B_1 \rightarrow B$ the eigensolution may also satisfy oddness relative to the $x$-axis (see Fig \ref{fig:prepert_orbs}). Similarly, at $N$ and along $N\Gamma$, the eigenstates transform under $\sigma_v$ as,
\beq
\hat{P}_{\sigma_v} \ket{\psi_{A_1, B_1}} = \pm  \ket{\psi_{A_1, B_1}}, \quad \hat{P}_{\sigma_v} \ket{\psi_{A, B}} = + \ket{\psi_{A, B}}.
\label{eq:sigma_v} \eeq This implies that the bands $(A_1, B_1)$ (at $N$) are compatible with $(A, A)$ (along $N\Gamma$). These compatibility relations are summarised pictorially in the unfurled IBZ path (Fig. \ref{fig:IBZ_path}(c)). Importantly, note that, in deriving equation \eqref{eq:effective_eqn} we have only assumed that $\bkappa_0$ belongs to a particular symmetry set band (surfaces $3-6$) (the band at $\bkappa_0$ must be continuously connected to the same band at $N$). Therefore, the compatibility relations allow us to choose any expansion point along the the path $\Gamma N X$ where the eigenfunction basis set, Eq. \eqref{eq:psi_expansion}, transforms accordingly i.e. $\ket{\psi_{A_1}} \rightarrow \ket{\psi_{A}}$.

\begin{table}[h!]
\begin{tabular}{|l|cc|c|}
\hline
\cellcolor[HTML]{EFEFEF}Classes $\rightarrow$ &                              &                                &                                                                              \\
\cellcolor[HTML]{EFEFEF}IR $\downarrow$       & \multirow{-2}{*}{$\sigma_h$} & \multirow{-2}{*}{$\sigma_{v}$} & \multirow{-2}{*}{\begin{tabular}[c]{@{}c@{}}Basis \\ functions\end{tabular}} \\ \hline
$A_1$                                         & $+1$                         & $+1$                           & $x^2, y^2, xy$                                                               \\
$B_1$                                         & $-1$                         & $+1$                           & $x, xy^2$                                                                    \\ \hline
\end{tabular}
\caption{$\sigma_{v, h}$ character table}
\label{table:c2_table}
\end{table}

\begin{table}[h!]
\begin{tabular}{|l|cc|c|}
\hline
\cellcolor[HTML]{EFEFEF}Classes $\rightarrow$ &                       &                                   &                                                                              \\
\cellcolor[HTML]{EFEFEF}IR $\downarrow$       & \multirow{-2}{*}{$E$} & \multirow{-2}{*}{$\sigma_{v, h}$} & \multirow{-2}{*}{\begin{tabular}[c]{@{}c@{}}Basis \\ functions\end{tabular}} \\ \hline
$A$                                           & $+1$                  & $+1$                              & $x^2, y^2, xy$                                                               \\
$B$                                           & $+1$                  & $-1$                              & $x, y, x^2y, xy^2$                                                           \\ \hline
\end{tabular}
\label{table:c2v_table_excerpt}
\caption{Selected portion of $C_{2v}$ character table.}
\end{table}

\begin{figure} [h!]
	\centering
		\includegraphics[width=0.45\textwidth]{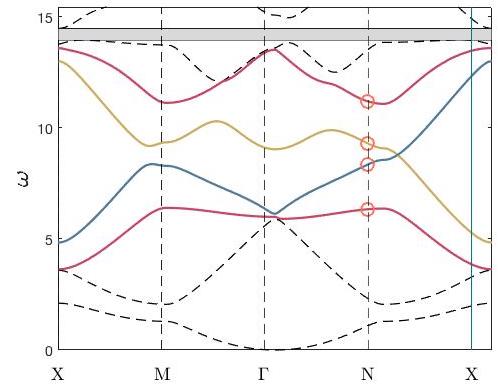}
	\caption{Dispersion curves $C_{2v}$ case (when $\omega_{B_1}>\omega_{A_1}$ at $N$) \textemdash Parameter values same as those in Fig. 2; $\sigma_v$ symmetry  present within physical space cell. In this instance, $B_1$ curve lies above $A_1$ curve at $N$, hence there is band crossing along $NX$.} 
\label{fig:dc_sse_dirac}
\end{figure}

In order to solve the 2-band eigenvalue problem, Eq. \eqref{eq:effective_eqn}, we compute the determinant of the truncated matrix,
\beq
\begin{vmatrix}
    -2\omega_{A} \Delta \omega_A \Lambda_A      &  \Delta\kappa_x \braket{\psi_{A}|\partial_x^3 + \partial_x\partial_y^2|\psi_{B}} \\
    \Delta\kappa_x \braket{\psi_{A}| \partial_x^3 + \partial_x\partial_y^2|\psi_{B}}^*      & -2\omega_{B} \Delta \omega_B \Lambda_B  
\end{vmatrix} = 0, 
\nonumber
\eeq where parity considerations \cite{dresselhaus_group_2008, heine_group_nodate} allows for simplification of the Hermitian matrix; the eigensolutions are evaluated at $\bkappa_0$. Solving the eigenvalue problem yields the following result,
\beq
2\omega_{A, B} \Delta \omega_{A, B} \Lambda_{A, B} = \pm    \Delta\kappa_x \braket{\psi_{A}| \partial_x^3 + \partial_x\partial_y^2|\psi_{B}},
\label{eq:local_linear}\eeq where the $\pm$ corresponds to the $A, B$ bands, respectively. This result implies that the $A, B$ bands have an identical slope, albeit with opposite gradients; hence, if, at $N$ an instance can be found where $\omega_{B_1} > \omega_{A_1}$  then the bands will invariably cross along the path $NX$. The parametric variation afforded to us, and encompassed in the variable $\Lambda_{A, B}$, merely increases or decreases the slope thereby increasing or decreasing the distance between $N$ and the Dirac point. 
Note that the Dirac cone occurs along the spatial symmetry path, $\sigma_h$, of the structure due to the opposite parities of the $A, B$ bands; band repulsion occurs along the $N \Gamma$ path 
[22] thereby resulting in a partial band gap along $N\Gamma$. If $\omega_{B_1} > \omega_{A_1}$, then the partial gap along $N\Gamma$ isolates the Dirac cone along a portion of the IBZ path, $\Gamma N X$.

\begin{figure} [h!]
	\centering
		\includegraphics[width=0.45\textwidth]{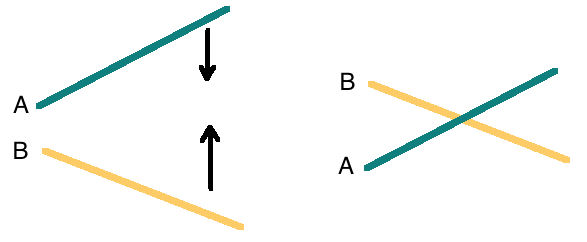}
	\caption{Effect of parametric tuning on $A, B$ bands \textemdash When $B$ curve lies about the $A$ curve, the parameters denoted by $\Lambda_{A, B}$ can be altered to change the intersection location. For our model, the number of masses, their location ($\sigma_v$ symmetry preserved) and their mass values can all be varied.} 
\label{fig:band_crossing}
\end{figure}

The distance between the Dirac and high-symmetry point is highly relevant for the transmission properties of the topological guide \cite{makwana_designing_2018}. \cite{makwana_designing_2018} stated that the transmission is better for short wave envelopes, as opposed to long wave envelopes, hence, for transmission post the nodal region, it is desirable to increase the distance between the Dirac cone and $N$. The latter is true due to the connection between the bulk and projected bandstructures \cite{bostan_design_2005}; the bulk BZ is reduced to a one-dimensional BZ because the only relevant wavevector component for a straight guide is the one parallel to the ZLM. All wavevectors are projected onto the $\Gamma M$ line in Fourier space, hence if the distance between $N$ and the Dirac cone is increased then the Fourier separation between oppositely propagating modes, along the topological guide, would be increased.
A mechanism to do this would be by altering the system parameters; Fig. \ref{fig:band_crossing} and Eq. \eqref{eq:local_linear} demonstrate that the slopes of the $A$ and $B$ bands can be increased or decreased by the system parameters thereby altering the position of the band intersection. 

\subsection{Breaking $\sigma_v$ symmetry}
\label{sec:C2v_bulk}

\begin{figure}
	\centering
	\begin{tabular}{c}
		(a) \\ \includegraphics[width=0.75\linewidth]{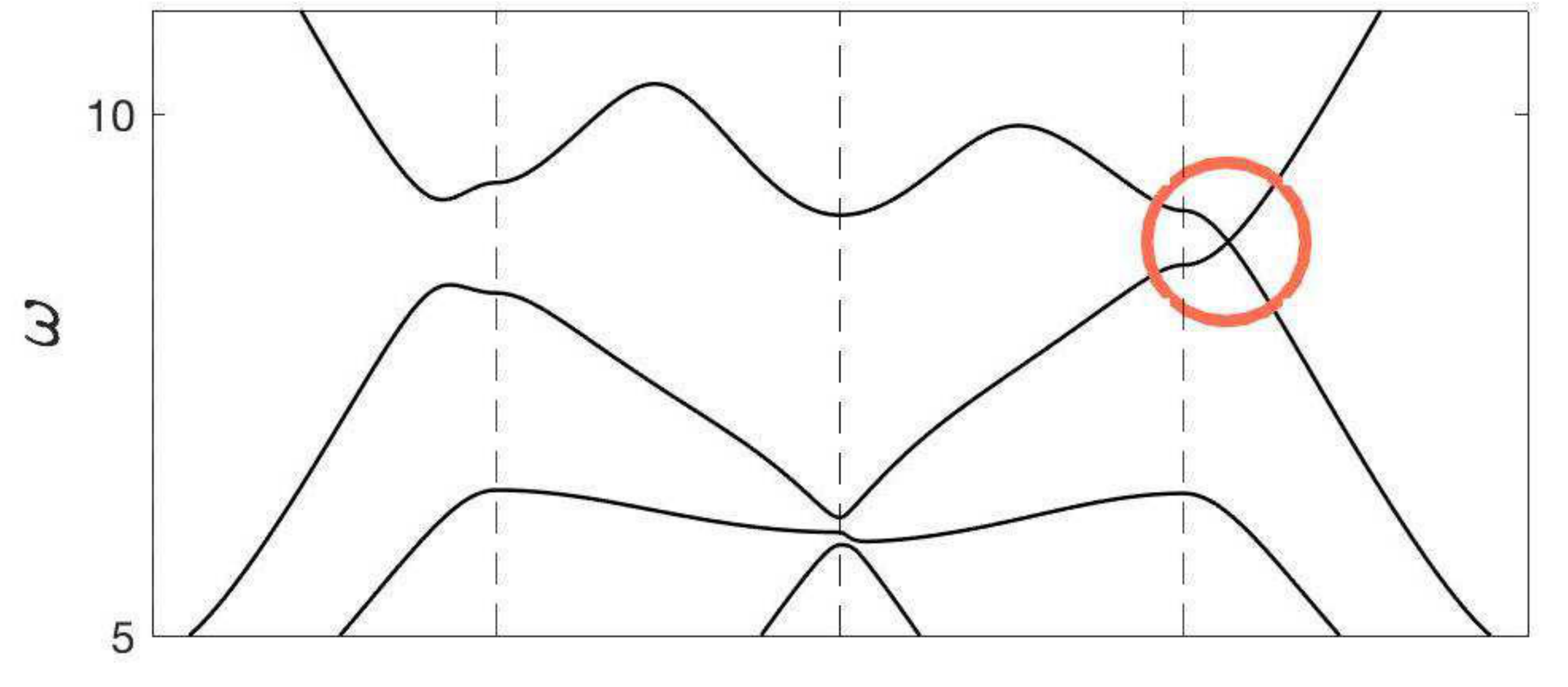} \\ (b) \\  \includegraphics[width=0.75\linewidth]{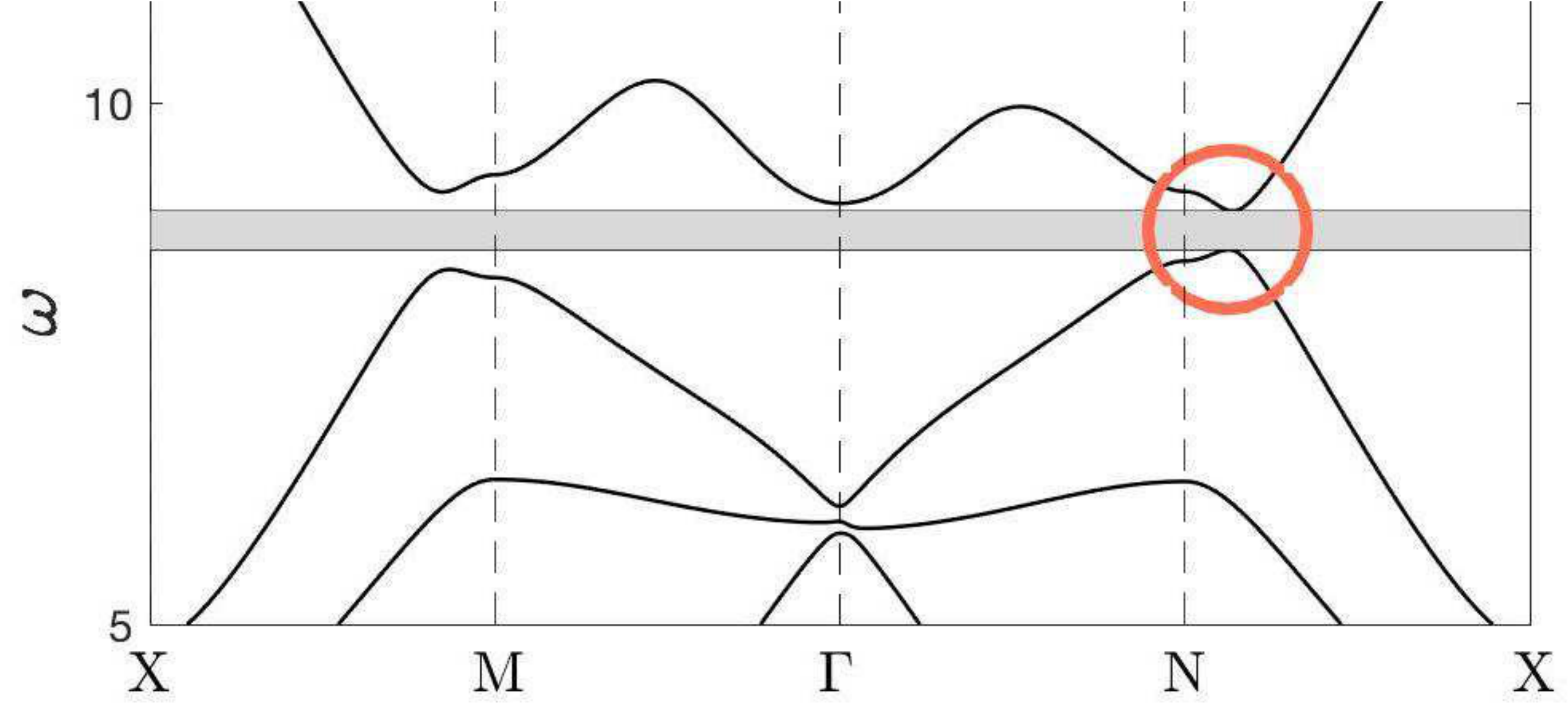}
	\end{tabular}
\caption{Dispersion curves for the $C_{2v}$ case. Panel (a) pre-perturbation curves, when structure in Fig. \ref{fig:C2v_case}(a) possesses $\sigma_v$ symmetry. Rotation of inclusion set in Fig. \ref{fig:C2v_case}(a) removes $\sigma_v$ symmetry and yields the post-perturbation curves shown in panel (b).}  \label{fig:C3v_curves}
\end{figure}

\begin{figure}[!htb]
    \centering
    \captionsetup{justification=raggedright}
     \begin{minipage}{0.5\textwidth}
        \hspace{-1.5cm}
         \includegraphics[width=0.55\linewidth]{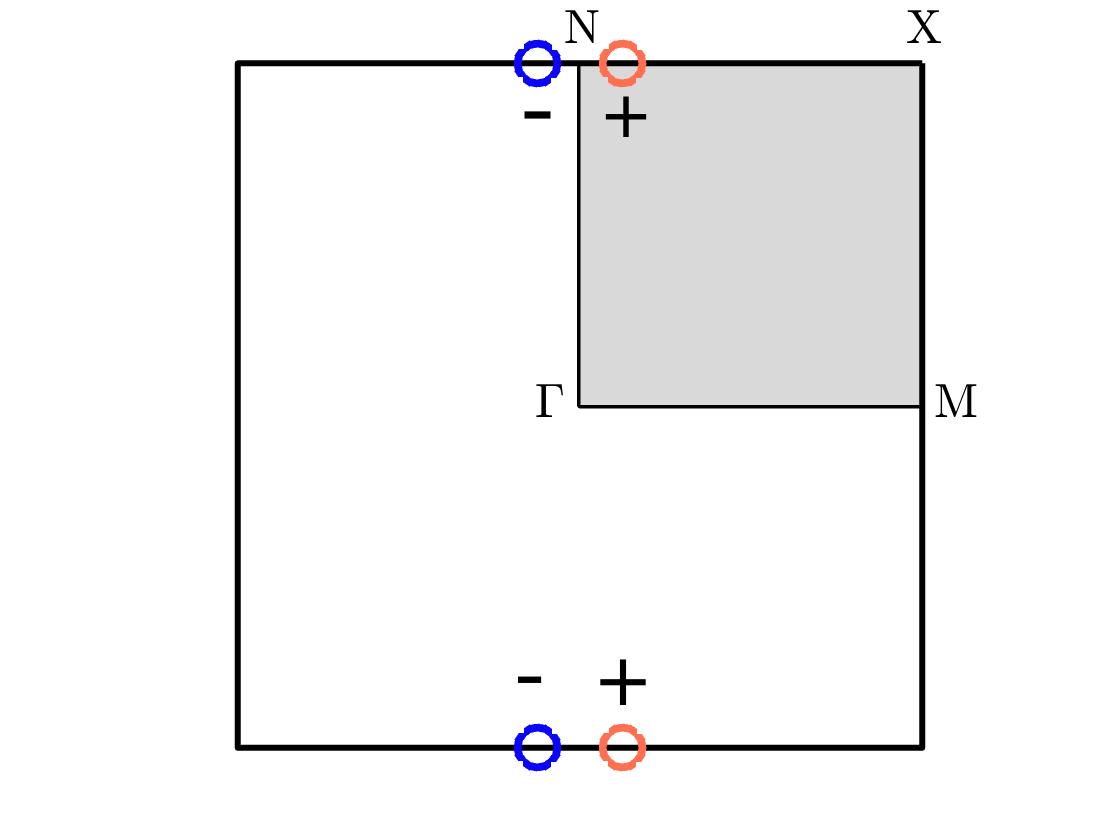}
        \label{fig:prob1_6_1}
    \end{minipage}
    \caption{IBZ (shaded region) within the Brillouin zone (BZ); circles indicate Dirac cone locations pre-perturbation, whilst $\pm$ denotes the signum of the post-perturbation Berry curvature. Dirac cones solely along single set of parallel HSLs, not both.}
    \label{fig:Berry_C2v}
\end{figure}

From the previous subsection we know that when the $\sigma_v$ symmetry is preserved an accidental Dirac degeneracy can be created; the bands coalescing along $NX$ in Fig. \ref{fig:C3v_curves}(a) are parametrically engineered to do so. An important nuance is that the Dirac points are solely located along the two high-symmetry lines (HSLs), parallel to $\sigma_h$, and not along the perpendicular HSLs (see Fig. \ref{fig:IBZ_path}(b)); this is critical when it comes to  energy-splitting. The $\sigma_v$ symmetry is lost in Fourier space when the internal set of inclusions is rotated and this breaks open the Dirac point to create a band-gap, Fig. \ref{fig:C3v_curves}(b).
The locally quadratic curves, in the vicinity of the former Dirac cones, carry nonzero Berry curvatures (Fig. \ref{fig:Berry_C2v}) which in turn leads to the generation of valley-Hall edge modes. Those regions on opposite sides of the $\sigma_v$ reflectional line carry Berry curvatures with opposite signum, Fig. \ref{fig:Berry_C2v}. In the next subsection, we shall show how, the locations of nonzero Berry curvatures, dictates how the geometrically distinct media are stacked.

\subsection{$C_{2v}$ adjoining ribbons}
\label{sec:C2v_ribbons}
\begin{figure} [h!]
\centering
	\includegraphics[scale=0.170]{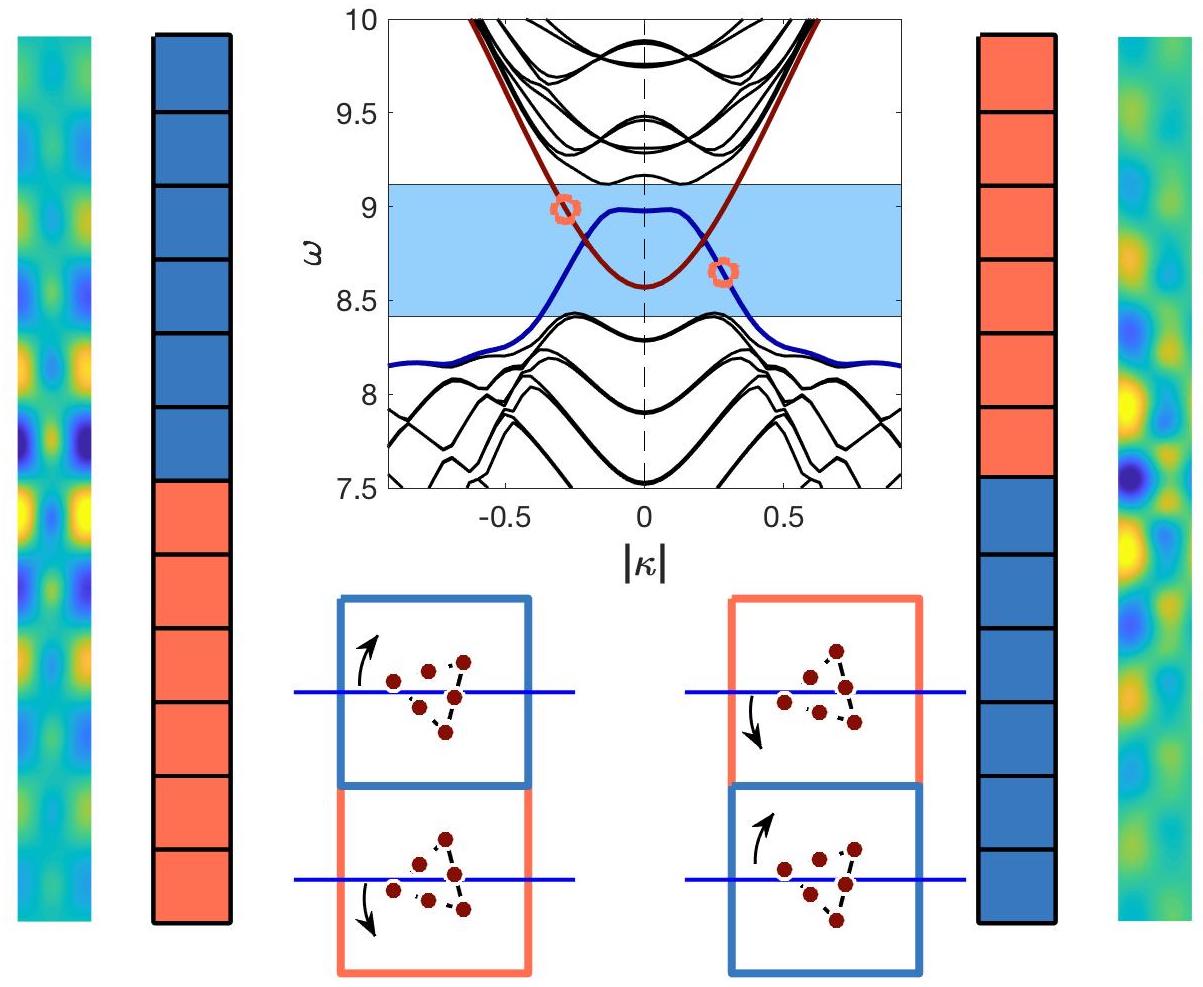}
\caption{Interfacial dispersion curves and ZLMs \textemdash \hspace{0.05cm} \emph{Top}: maroon curve arises when blue medium stacked over orange (left-sided ribbons), whilst navy curves, when orange over blue (right-sided ribbons). \emph{Left}: even-parity ZLM, $\omega = 9.00$. \emph{Right}: odd-parity ZLM, $\omega = 8.60$.}
\label{fig:C3v_edge_modes}
\end{figure}

Attaching two topological media, with opposite Berry curvatures \cite{xiao_valley-contrasting_2007} yields broadband chiral edge states. This is achieved by placing one gapped medium, above its $\sigma_v$ reflected twin; in essence, the stacking in Fourier space results in regions of opposite Berry curvatures overlaying each other, this local disparity ensures the presence of valley-Hall edge modes. The two distinct orderings of the media create two distinct interfaces, as seen in Fig. \ref{fig:C3v_edge_modes} one of which supports only the even modes and the other only the odd modes. This evenness and oddness of the edge modes is inherited from the even and odd bulk modes, Fig. \ref{fig:prepert_orbs}. The \emph{gapless} curves are a symptom of the topologically nontrivial nature of the edge states; this is akin to the valley-Hall modes seen for the zigzag interface within hexagonal structures.

The simplicity of this construction, the apriori knowledge of how to tessellate the two media to produce these broadband edge states, and the added robustness \cite{makwana_designing_2018} are the main benefits of these topological valley-Hall modes.  The additional functionality of having a three-way topological splitter (Fig. \ref{fig:3_way_splitter}) comes with a caveat: The Fourier separation between the valleys controls the intervalley scattering and the smaller separation in the square lattice, Fig. \ref{fig:C3v_edge_modes}, vis-a-vis that for graphene-like structures \cite{makwana_designing_2018} leads to increased scattering. This can be mitigated as the Fourier separation can be artificially increased by parametrically increasing the distance between the Dirac cone and $N$ in Fig. \ref{fig:C3v_curves} thereby acting to increase the robustness of the edge states against shorter-range defects.

\subsection{$C_{2v}$ ZLM and absence of post-bend propagation}
\label{sec:C2v_scattering}
\begin{figure}
	\centering
	\begin{tabular}{cc}
		(a) & (b) \\
		\includegraphics[height=0.40\linewidth]{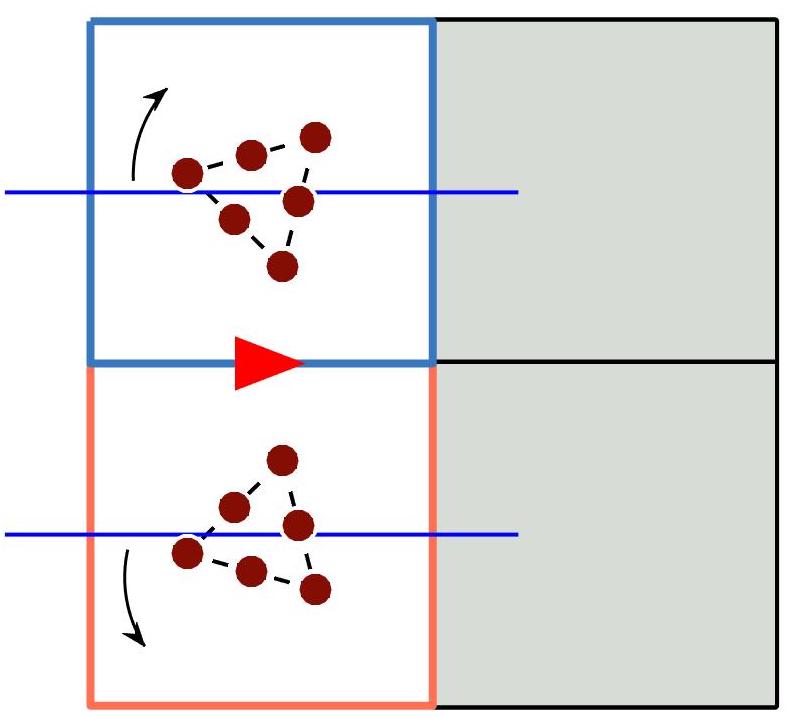} &
		\includegraphics[width=0.40\linewidth]{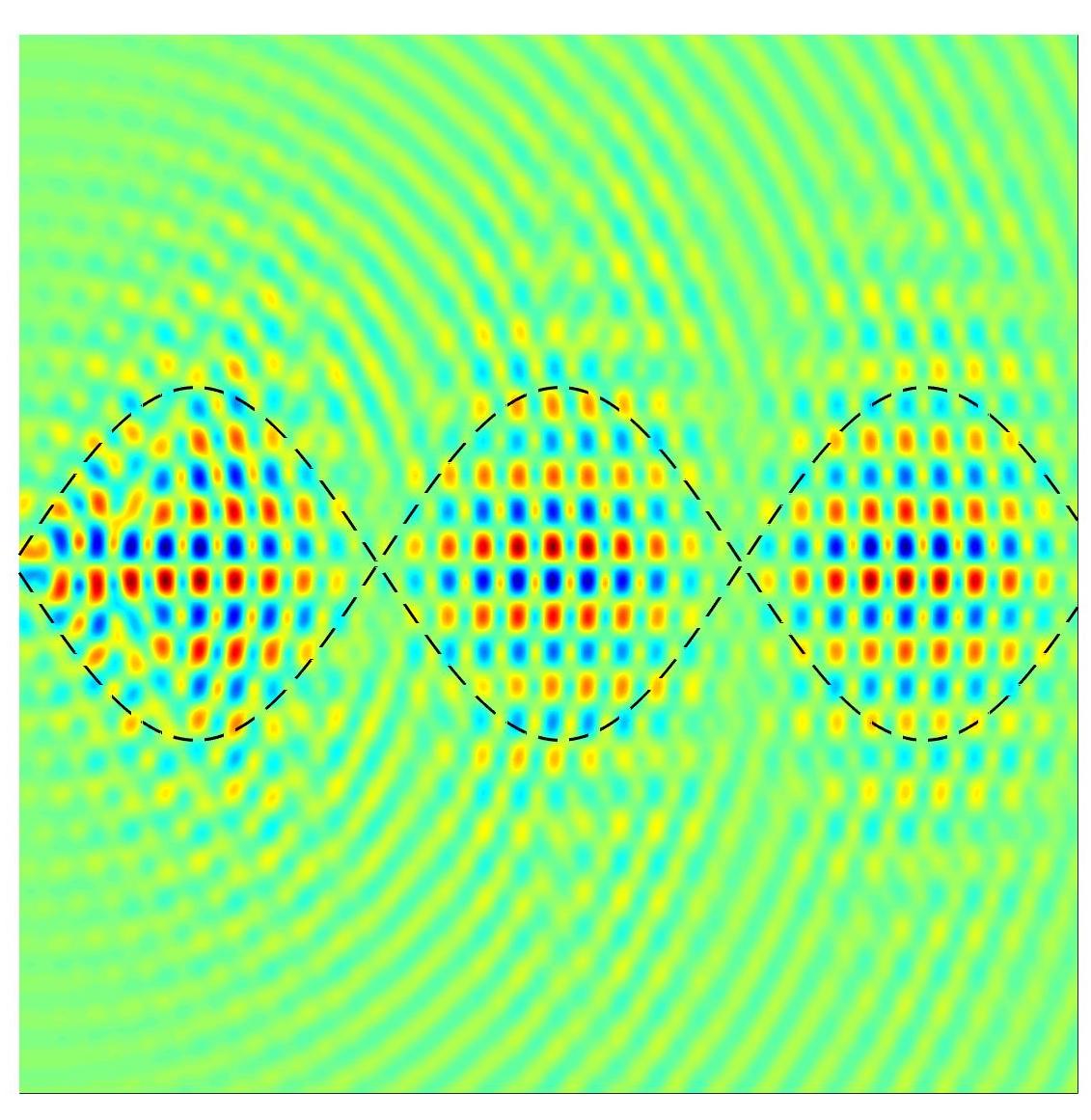}
	\end{tabular}
\caption{Odd-parity ZLM \textemdash (a) Odd-parity mode along blue over orange interface; importantly, there is no arrangement for the grey cells that ensures a vertical mode. As there are no well-defined valleys of nonzero Berry curvature along the vertical edges of the BZ in Fig \ref{fig:Berry_C2v}, energy cannot be steered around a $\pi/2$ bend. (b) Dipolar source placed at leftmost edge, excites odd-parity ZLM. The periodicity of the long-scale envelope is clearly evident; outline around envelope derived from HFH (Appendix). Backscattering can be minimised via parametric variation (by decreasing the wavelength of the energy-carrying envelope). }
\label{fig:C2v_ZLM}
\end{figure}

The property of the $C_{2v}$ case that prohibits propagation around the bend is the absence of  well-defined valleys with nonzero Berry curvature, along the vertical HSLs of the BZ, see Fig. \ref{fig:Berry_C2v}. Hence, there is no arrangement that can be placed to the right of either stacking in Fig. \ref{fig:C3v_edge_modes} to obtain a ZLM perpendicular to the blue-orange interface, Fig. \ref{fig:C2v_ZLM}. The ZLM has a long-scale periodic envelope that can be captured using an effective medium theory \cite{makwana_wave_2016} (Appendix).  Knowledge of the long-scale envelope is especially useful for these finite length interfaces as it can used to minimise the backscattering as one has, in effect, a Fabry-P\'erot resonator.

To summarise, for this $C_{2v}$ case, there are ZLMs along straight interfaces, however the energy cannot navigate around a $\pi/2$ bend because there is no post-bend mode to couple with. 


\section{$C_{4v}$ cellular structure}
\label{sec:C4v}
We now extend the concepts illustrated in Sec. \ref{sec:C2v} to a cellular structure that possesses $C_{4v}$ point group symmetry at $\Gamma$ (see Fig. \ref{fig:C4v_curves}(a)). Due to this structure possessing two perpendicular mirror symmetries, as opposed to one, there exists regions of nonzero Berry curvature along both edges of the square BZ, subsection \ref{sec:C4v_bulk} (Fig. \ref{fig:C4v_curves}(b)); this will be shown to yield propagation around a corner (subsection \ref{sec:bend}) aswell as a three-way splitting of energy (subsection \ref{sec:three_way}). An extensive comparison between the square structures, discussed within this article, and the earlier valleytronics models based upon graphene-like structures \cite{makwana_designing_2018, cha_experimental_2018, he_acoustic_2016, he_silicon--insulator_2018, he_two-dimensional_2019, khanikaev_two-dimensional_2017, nanthakumar_inverse_2019, ozawa_topological_2019, qiao_current_2014, schomerus_helical_2010, shen_valley-projected_2019, xia_topologically_2019, yan_-chip_2018, ye_observation_2017, cheng_robust_2016, wu_direct_2017, xia_topological_2017, zhang_manipulation_2018, qiao_electronic_2011} will be pictorially shown at the end of subsection \ref{sec:three_way}.

\subsection{Breaking $\sigma_{v, h}$ symmetries}
\label{sec:C4v_bulk}
\begin{figure} [h!]
	\begin{tabular}{cc}
		(a) & (b) \\
        \includegraphics[height=0.45\linewidth]{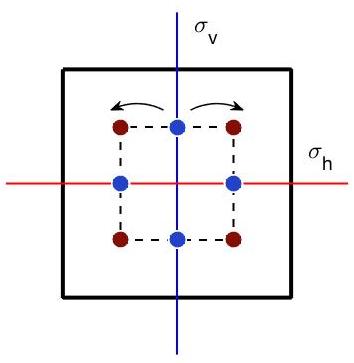} & 
         \includegraphics[height=0.40\linewidth]{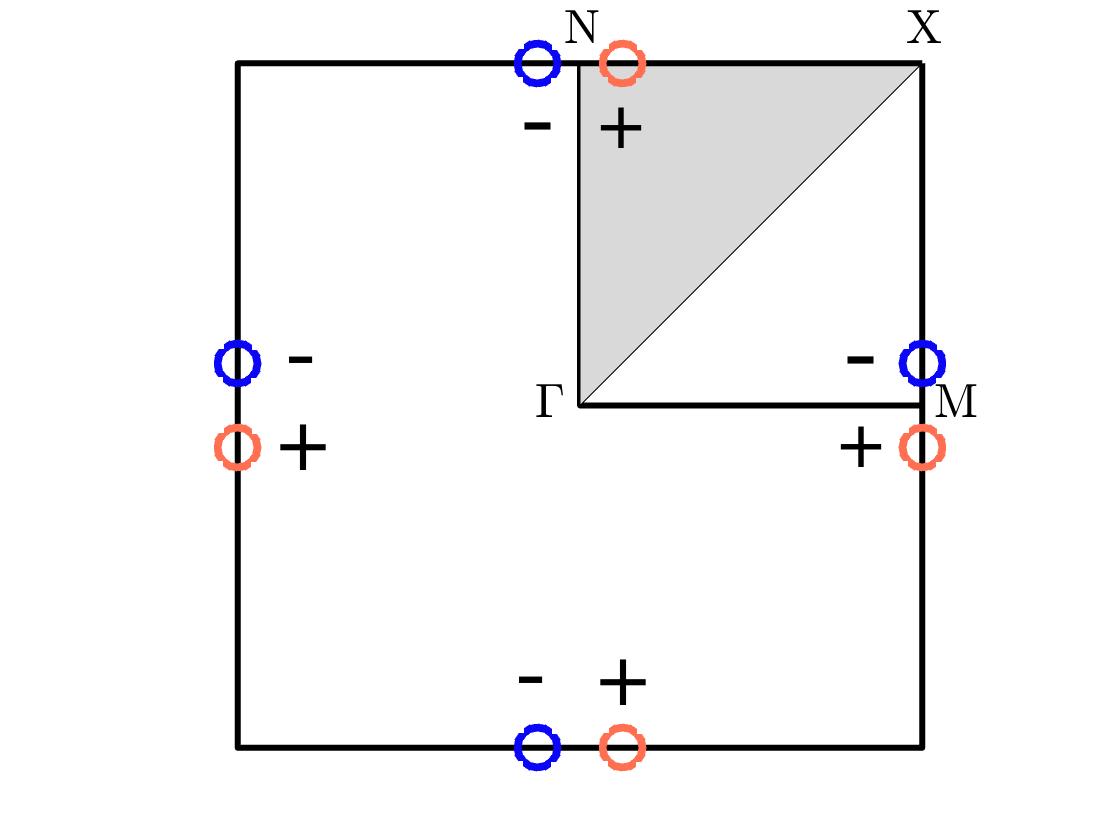} \\
        \multicolumn{2}{c}{(c)} \\
        \multicolumn{2}{c}{\includegraphics[width=0.70\linewidth]{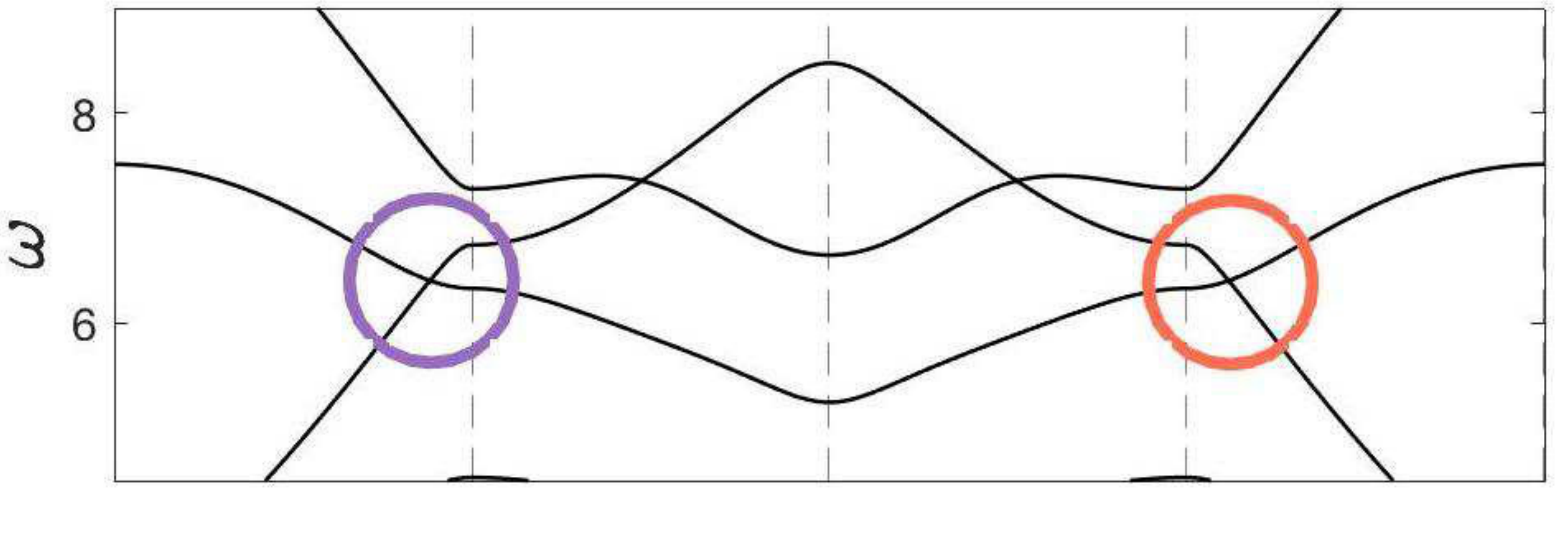}} \\
        \multicolumn{2}{c}{(d)} \\
        \multicolumn{2}{c}{\includegraphics[width=0.70\linewidth]{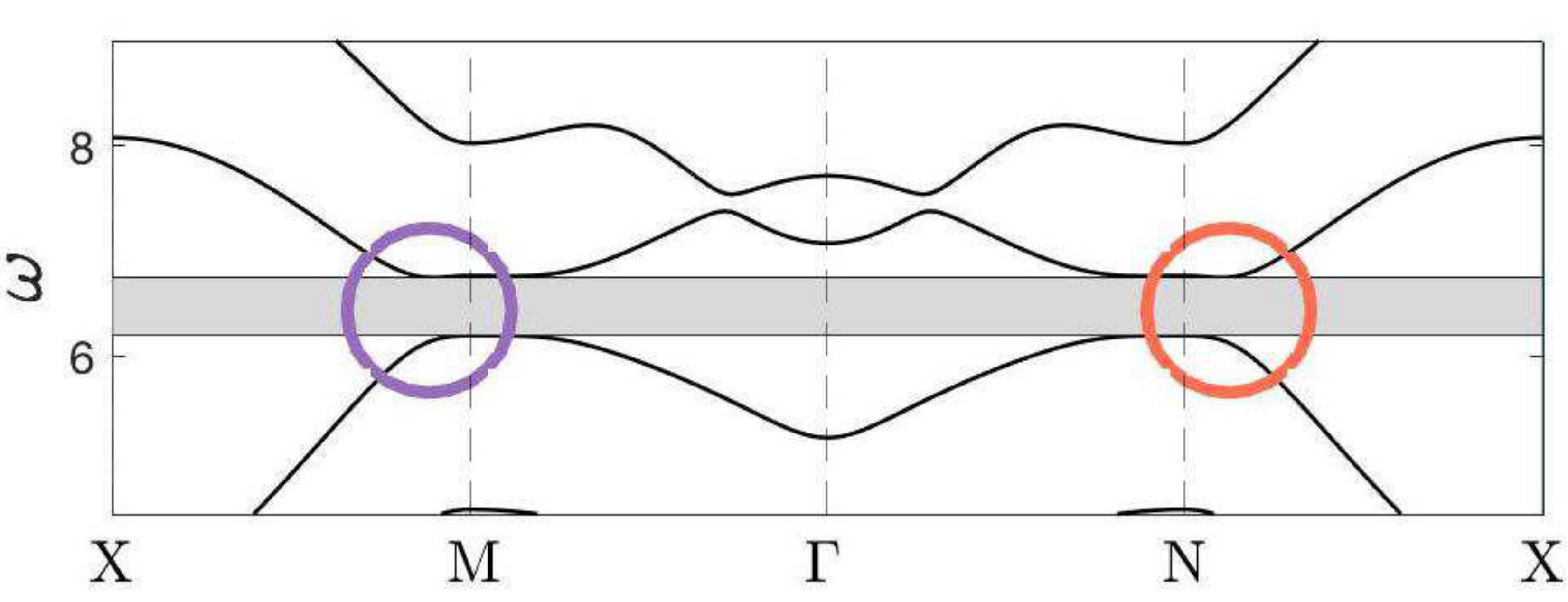}}     
	\end{tabular}
\caption{Dispersion curves $C_{4v}$ case \textemdash(a) cellular structure shown; maroon mass value of $1$, blue mass value of $2$, lattice constant of $2$, centroid to vertex mass distance  of $0.70$. Pre-perturbation structure has $\sigma_v$ and $\sigma_h$ symmetries, both of these symmetries are broken in the post-perturbation structure.  (b) shows IBZ (shaded region) within the BZ; circles indicate Dirac cone locations pre-perturbation, whilst $\pm$ denotes the signum of the Berry curvature, post-perturbation. Unlike the $C_{2v}$ case (Fig. \ref{fig:C3v_curves}), Dirac cones now present along both sets of parallel HSLs. (c) pre-perturbation dispersion curves. We have opted to plot around the IBZ of the $C_{2v}$ case (Fig. \ref{fig:C3v_curves}) in order to explicitly show the Dirac cone that arises from the added $\sigma_h$ symmetry. (d) post-perturbation dispersion curves. If we were to plot along the $C_{4v}$ IBZ an identical band-gap, in location and width, would be present.}
\label{fig:C4v_curves}
\end{figure}

The $C_{4v}$ case, Fig. \ref{fig:C4v_curves}(a), is reminiscent of the $C_{2v}$ case but now with the addition of $\sigma_h$ symmetry in physical space. This reflectional symmetry yields additional Dirac cones along, a parallel set of HSLs, perpendicular to those connected with the $\sigma_v$ symmetry (Fig. \ref{fig:C4v_curves}(b)). This is evident, for the unperturbed $C_{4v}$ case, in its dispersion curves, Fig. \ref{fig:C4v_curves}(c); note that we have plotted around the $C_{2v}$ IBZ to clearly illustrate the correspondence between the two sets of dispersion curves, Fig. \ref{fig:C3v_curves}(a) and Fig. \ref{fig:C4v_curves}(c). The additional Dirac cone, for the $C_{4v}$ case, along $XM$ is due to the additional $\sigma_h$ reflectional symmetry in physical space. Rotating the inclusion set (Fig. \ref{fig:C4v_curves}(a)) results in the breaking of both $\sigma_{h, v}$ symmetries thereby opening up a band-gap (Fig. \ref{fig:C4v_curves}(d)). Importantly, an identical band-gap is present whether we're plotting along the $C_{2v}$ or $C_{4v}$ IBZ's.

In the subsequent section, we demonstrate how the additional reflectional symmetry enables mode coupling from the pre-bend to post-bend ZLM thereby allowing for energy navigation around a corner.

\subsection{Propagation around a bend}
\label{sec:bend}
\subsubsection*{$C_{4v}$ adjoining ribbons and comparison with $C_{2v}$ case}
\label{sec:C4v_ribbons}
\begin{figure}
	\begin{tabular}{cc}
		(a) & (b) \\
		\includegraphics[width=0.175\linewidth]{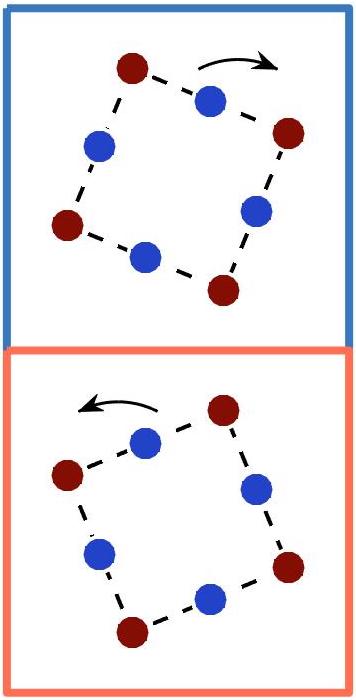} &
		\includegraphics[width=0.325\linewidth]{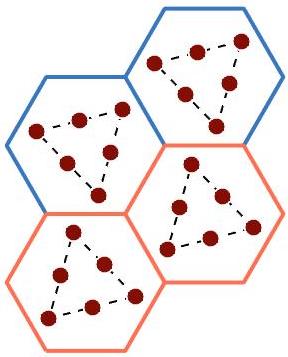} \\
		(c) & (d) \\
		\includegraphics[width=0.175\linewidth]{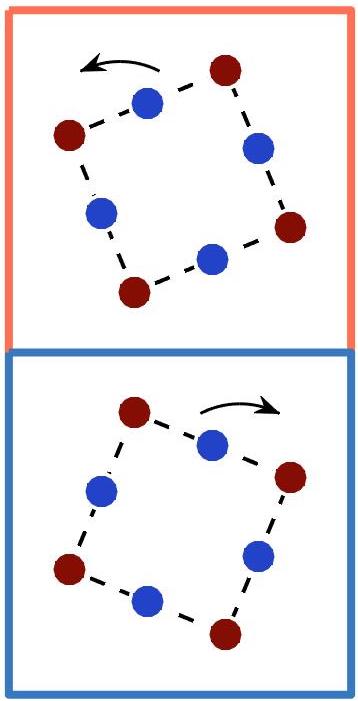} &
		\includegraphics[width=0.425\linewidth]{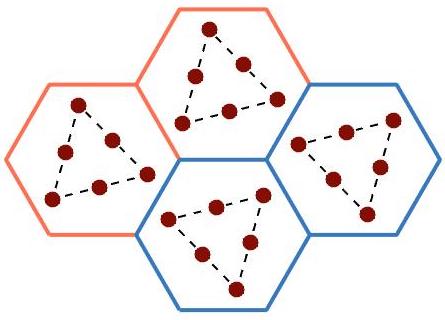}
	\end{tabular}
	\caption{Interface comparison between $C_{4v}$ case (a, c) and graphene-like structure (b, d) \textemdash representative hexagonal structure taken from \cite{makwana_designing_2018}. Evidently, the two hexagonal zigzag interfaces that host ZLMs are distinct whilst, the two square interfaces, are identical under TRS. Even and odd-parity edge modes exist along the \emph{same} interface for the $C_{4v}$ case and \emph{different} interfaces for the graphene-like structures and $C_{2v}$ cases (Fig. \ref{fig:C3v_edge_modes}). This latter point is what allows for coupling between the pre-bend and post-bend modes, Fig. \ref{fig:Schem_Bend}, for the $C_{4v}$ case but not the $C_{2v}$. Crucially, this property is also what yields three-way splitting for the $C_{4v}$ case, Fig. \ref{fig:3_way_splitter}, but \emph{not} for the graphene-like structures.}
	\label{fig:interfaces}
\end{figure}

A crucial property that allows for wave steering for the $C_{4v}$ case is the presence of Dirac cones along both edges of the BZ. Another important property is that, like the $C_{2v}$ case, both, even and odd edge modes exist, however they are now present along the \emph{same} interface as opposed to different interfaces. The orthogonality of these opposite-parity modes ensures that they do not couple along the same edge. The presence of both parity modes along the same interface (for the $C_{4v}$ case) arises from 
the relationship between the orange over blue stacking and its reverse (Figs. \ref{fig:interfaces}(a), (c)). Specifically, it is clearly evident from Figs. \ref{fig:interfaces}(a), (c) that a right propagating mode for one stacking is a left propagating mode on the other and vice versa. This special property is also what allows for the three-way splitting of energy (see subsection \ref{sec:three_way}).

\begin{figure}
	\includegraphics[width=0.5\textwidth]{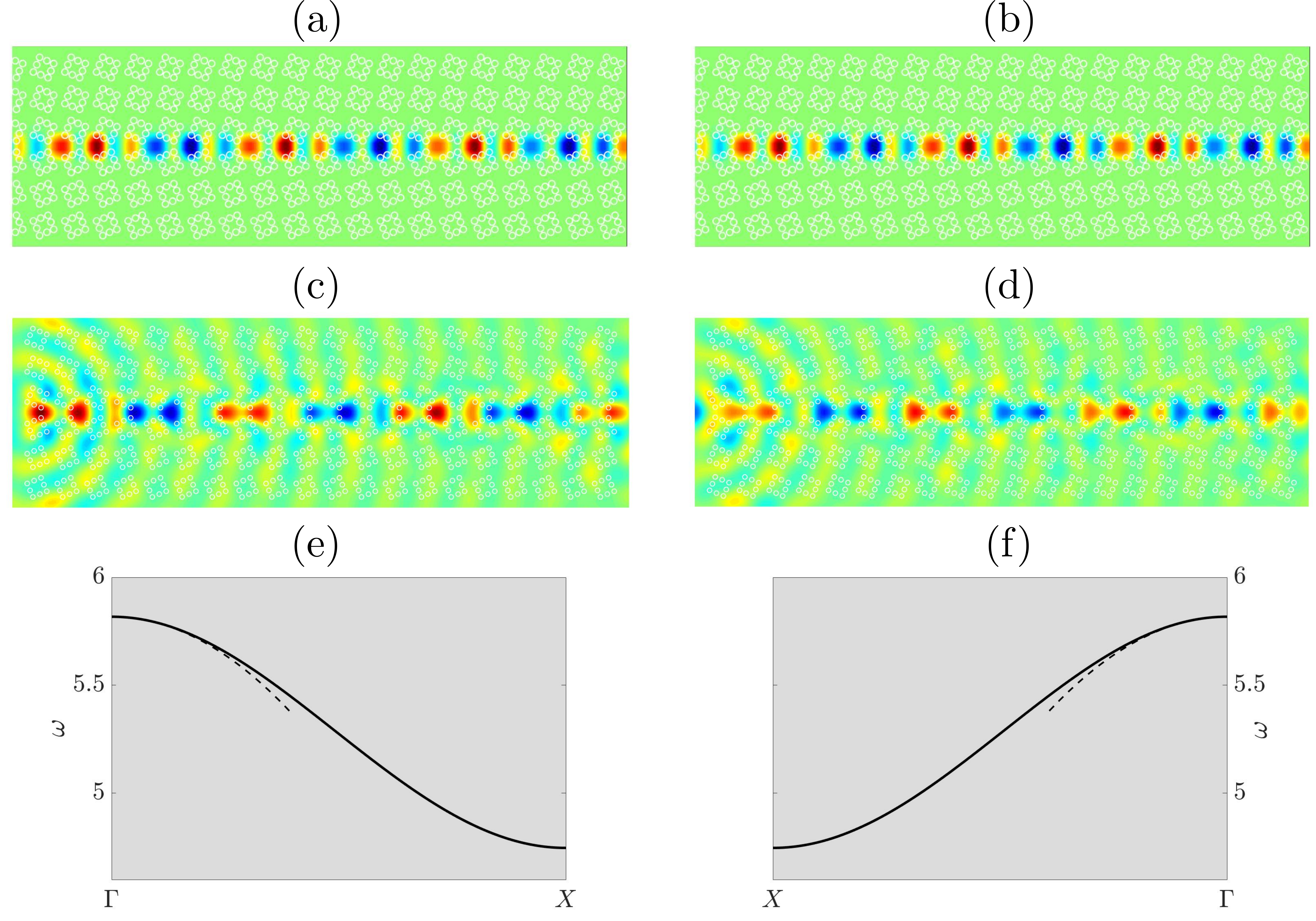}
	\caption{$C_{4v}$ even-parity interfacial mode \textemdash \emph{Left} and \emph{right} columns pertain to the orange over blue and blue over orange stackings, respectively. (a-b) Eigensolutions obtained from Fourier-Hermite method. (c-d) scattering solutions. (e-f) even-parity interfacial curve, for both stackings, are identical, HFH asymptotics \cite{chaplain_rayleigh-bloch_2018} also shown (dashed lines).}
	\label{fig:C4v_edge_modes}
\end{figure}

We now move onto deriving an edge mode for the $C_{4v}$ case. Due to there being only a single unique interface, we choose to use a Fourier-Hermite spectral method \cite{chaplain_rayleigh-bloch_2018}, that purely finds the decaying solution along a single interface, as opposed to simultaneously along both; the latter occurs when the PWE method is used in conjunction with two-dimensional periodic Bloch conditions. 
Hence, from Fig. \ref{fig:C4v_edge_modes}, we clearly see that, for a variant of the  $C_{4v}$ case, the orange over blue (Figs. \ref{fig:C4v_edge_modes} (a), (c) and (e)) or blue over orange (Figs. \ref{fig:C4v_edge_modes}(b), (d) and (f)) stacking  yields an even-parity decaying mode. More specifically, the orange over blue stacking gives solutions to the right of $\Gamma$ whilst the blue over orange yields solutions to the left of $\Gamma$; this implies that the two stackings host the same mode and are TRS pairs of each other. Note that a parametric variant of the $C_{4v}$ case was used to ensure faster convergence of the Fourier-Hermite spectral method. The local curvature, and thereby the characterisation of the envelope, is obtained for modes in the vicinity of $\Gamma$ (see asymptotics in Fig. \ref{fig:C4v_edge_modes}(e) and (f)). Similar to the earlier $C_{2v}$ structure the edge states that arise are topologically nontrivial and gapless \cite{xia_observation_2018}.

\subsubsection*{Transmission around a bend}
\begin{figure}
	\begin{tabular}{cc}
		(a) & (b) \\
		\includegraphics[height=0.18\textwidth]{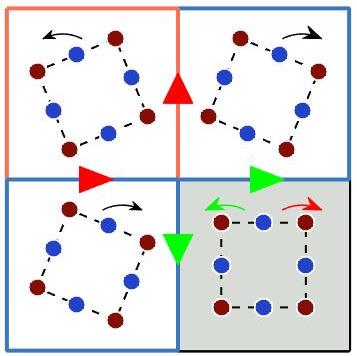} &
		\includegraphics[height=0.18\textwidth]{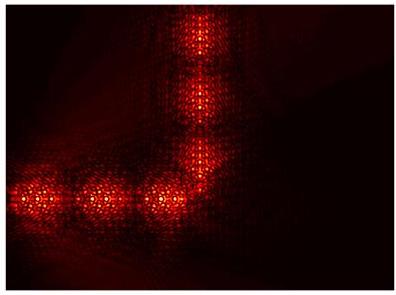}
	\end{tabular}
\caption{Wave-steering and energy-splitting \textemdash (a) by extending this nodal region outwards the entire structured domain for both effects is obtained. If the bottom-right quadrants inclusion set is rotated rightwards then a left-sided incident ZLM would follow the red arrows around the bend; leading to the modal pattern in the right panel. If the same set of inclusions is rotated leftwards, then energy is partitioned three-ways away from the nodal point, yielding the three-way energy-splitter, Fig. \ref{fig:3_way_splitter}. (b) example of topological wave steering. Similar to $C_{2v}$ ZLM, Fig. \ref{fig:C2v_ZLM}(b), long-scale modulation is distinguishable from the short-scale oscillations. }
\label{fig:Schem_Bend}
\end{figure}

The perturbed $C_{4v}$ system has valleys of nonzero Berry curvatures along all HSLs of the BZ (Fig. \ref{fig:C4v_curves}(b)). This allows for the strategic arrangement of four structured media such that valleys of opposite Berry curvature overlay each other along, both, horizontal and vertical interfaces, Fig \ref{fig:Schem_Bend}(a). This strategic arrangement necessitates the existence of broadband ZLMs along both of these interfaces simultaneously; therefore, unlike the $C_{2v}$ case, energy is navigable around bends.  

The four-cell arrangement shown in Fig. \ref{fig:Schem_Bend}(a) encompasses the design of the nodal region (and by extending it outwards, the entire region) for the $\pi/2$ wave steerer and three-way energy-splitter. If the bottom-right inclusion set is rotated clockwise then a wave incident along the leftmost interface will follow the red arrows around the $\pi/2$ bend. The indistinguishable, pre- and post-bend interfaces, ensure that, as the energy traverses the turning point, an even-parity mode will couple into itself. An example of, topological wave steering around a bend, is shown in Fig. \ref{fig:Schem_Bend}(b). Notably, the $\pi/2$  wave steerers observed within hexagonal structures require coupling between a zigzag mode with an armchair mode. The latter termination hosts topologically \emph{trivial} edge states due to the overlaying regions of identical Berry curvature resulting in gapped states. Contrastingly, the structure shown in Fig. \ref{fig:Schem_Bend} allows for topologically \emph{nontrivial} $\pi/2$ wave steering.

Similar to the $C_{2v}$ ZLM, the short-scale oscillations are discernible from the long-scale modulation. The importance of this long-scale modulation is numerically elucidated in the subsequent section.

\subsubsection*{Relevance of envelope to transmission around a bend}
\label{sec:C4v_envelope}

\begin{figure}
	\begin{tabular}{cc}
		(a) & (b) \\
		\includegraphics[width=0.45\linewidth]{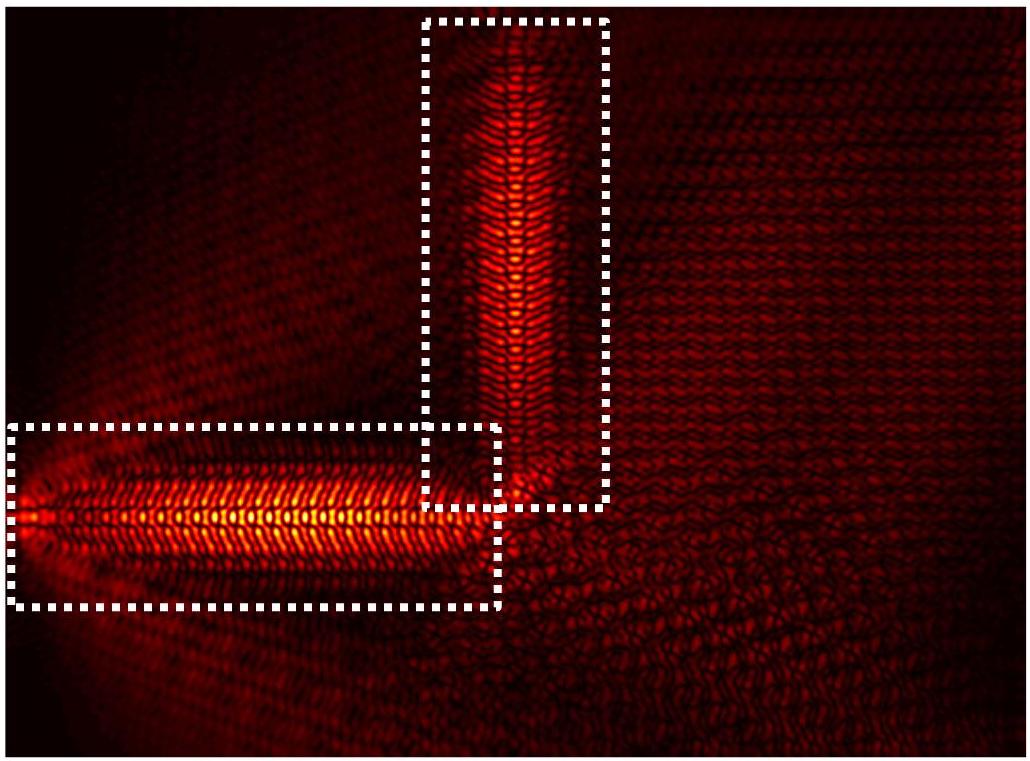} &
		\includegraphics[width=0.45\linewidth]{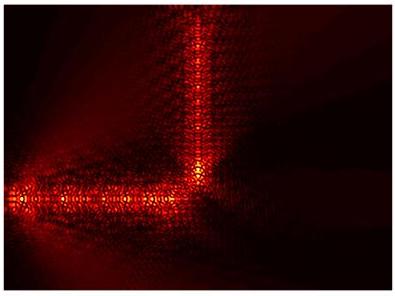} \\
		(c) & (d) \\
		\includegraphics[width=0.45\linewidth]{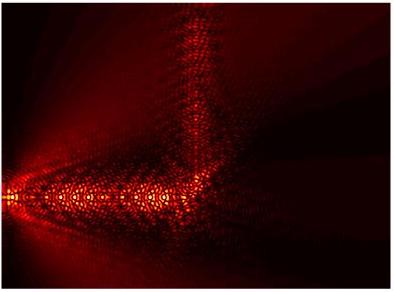} &
		\includegraphics[width=0.45\linewidth]{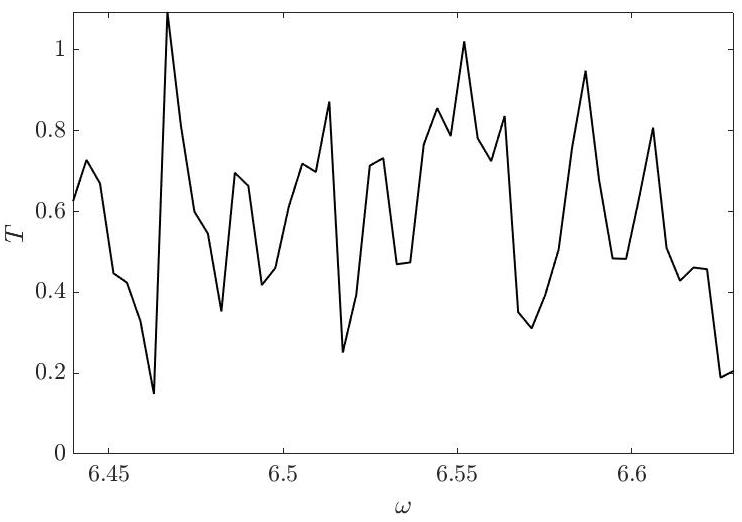}
	\end{tabular}
	\caption{Wave steering examples \textemdash (a-b) panels show different examples of high-transmission wave steering. Notably, in each of these cases the long-scale envelope is discernible, and more importantly, the wavelengths of these envelopes is entirely contained within the first lead thereby allowing for near-perfect transmission around the bend. (c) shows an instance where the the incident ZLM impacts the turning point with maximum amplitude, resulting in significant backscattering. Right-sided panel shows the highly variable transmission of this long-wavelength wave steerer. Transmission is calculated from the ratio of the intensities contained within the two boxes (shown in the upper-left panel). The overlap of the boxes introduces a small numerical error that can yield unphysical transmissions (e.g. $\omega = 6.465$)}
	\label{fig:many_bends}
\end{figure}

The characterisation of the energy-carrying envelope is important, as the tuning of it can lead to higher transmission along finite length interfaces. This principle is elucidated by examining the wave-steering example, Fig. \ref{fig:many_bends}; using finite element integration the intensity of the wave-field in each arm of such a steerer is calculated. The ratio of these intensities is the measure of the transmission of the wave steerer (Fig. \ref{fig:many_bends}(d)); this quantity can be seen to oscillate rapidly across the band-gap. This is similar to the behaviour of conventional Fabry-P\'erot resonators, where for maximal transmission an integer number of wavelengths must be completely contained in each lead. Thus the length of the interfaces is of importance for optimising the transmission. This effect is clearly seen by the contrast in transmission (in Fig. \ref{fig:many_bends}) between Fig. \ref{fig:many_bends}(c) and (a-b). Despite the paradigm utilising the valley-Hall topological phase, the robustness and bandwidth of the effect can be further increased, by parametic variation, introducing a TRS-breaking active component, nonlinearity and/or resonators within the nodal region.

\subsection{Topological 3-way splitter}
\label{sec:three_way}
We now move onto the construction of the three-way energy-splitter; rotating the bottom-right inclusion set anti-clockwise, in Fig. \ref{fig:Schem_Bend}(a); results in four partitions of geometrically distinct media. A wave incoming, from the leftmost interface, will now follow, both, the red and green arrows thereby splitting the energy three-ways. The resulting scattering solution, for a monopolar source, is shown in Fig. \ref{fig:3_way_splitter} ; the topological nature of the modes is demonstrated by the chiral fluxes. The three-way splitter can be tuned to a wave steerer by rotating the cellular structures in lower-right quadrant.

\begin{figure} [h!]
	\begin{tabular}{cc}
		(a) & (b) \\
		\includegraphics[width=0.515\linewidth]{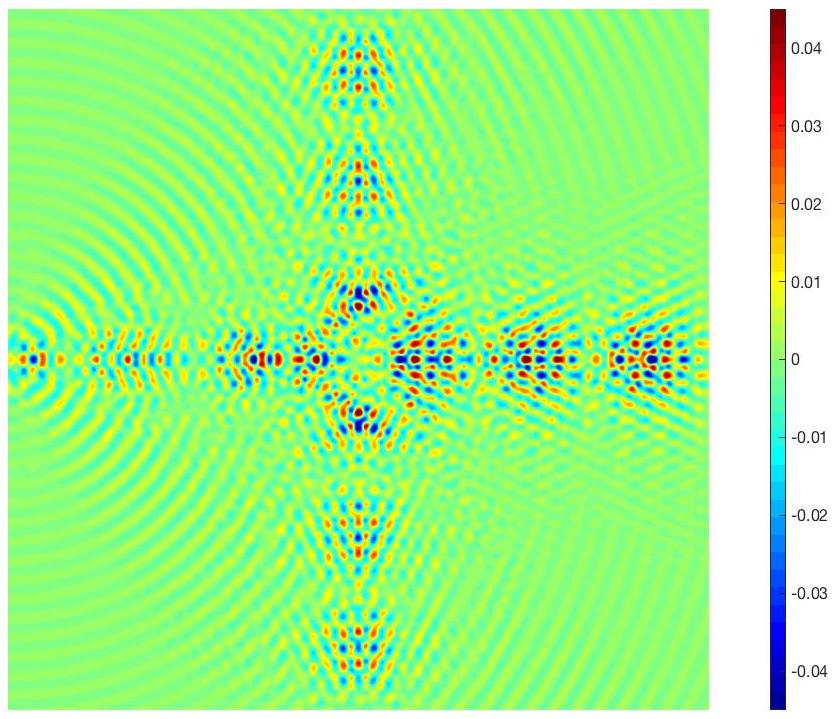} &
		\includegraphics[width=0.445\linewidth]{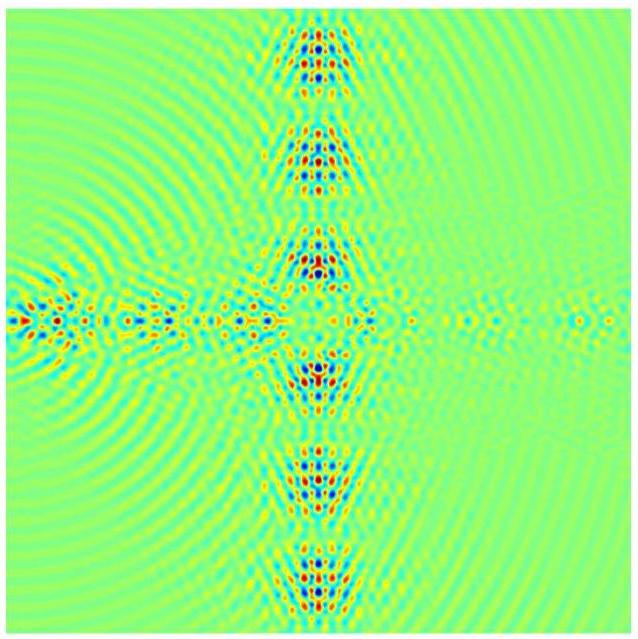} 
	\end{tabular}
\caption{Real (a) and imaginary (b) components of the displacement field shown in Fig. \ref{fig:3_way_splitter}. Notably in panel (a), the monopolar source triggers an even-parity ZLM along the left-sided interface; this mode couples into identical parity modes along all three outgoing leads. The absence of excitation along the right-hand lead, in panel (b), indicates that there is a phase difference between this lead and the other three excited leads.}
\label{fig:real_imag_splitter}
\end{figure}

For a mode to couple, from one lead to another, the chirality and the $\bkappa$'s of the modes must match \cite{makwana_designing_2018, cha_experimental_2018, he_acoustic_2016, he_silicon--insulator_2018, he_two-dimensional_2019, khanikaev_two-dimensional_2017, nanthakumar_inverse_2019, ozawa_topological_2019, qiao_current_2014, schomerus_helical_2010, shen_valley-projected_2019, xia_topologically_2019, yan_-chip_2018, ye_observation_2017, cheng_robust_2016, wu_direct_2017, xia_topological_2017, zhang_manipulation_2018, qiao_electronic_2011} (the $\bkappa$'s may be $2\pi/a$, phase shifts of each other, where $a$ is the lattice constant by periodicity considerations). For the square $C_{4v}$ case this condition is satisfied due to the relationship between the interfaces; an incident even mode couples to itself along the three exit leads, Fig. \ref{fig:real_imag_splitter}(a). The interfaces for the upper and lower leads (Fig. \ref{fig:3_way_splitter}) matches the left lead interface (orange over blue),  hence an incident even mode at valley $K$ will couple to itself along the upper and lower leads. \emph{Importantly}, the right-sided interface (blue over orange) is the reverse of the left-sided interface, hence a right propagating mode on the right-sided interface is identical to a left propagating mode on the left-sided interface; therefore, for seamless coupling between these two leads, an incident $K$ mode needs to couple into an outgoing $K'$ mode (where $K' = K \pm 2\pi/a$) with matching chirality. This phase difference, between, the right-hand lead and the other three leads, is evident when we plot the imaginary component of the three-way energy-splitter (Fig. \ref{fig:real_imag_splitter}(b)). This additional phase of $2\pi/a$ is acquired in a similar manner to the  $2\pi/a$ phase obtained when an incident wave passes through a gratings coupler \cite{shun_lien_chuang_physics_2009}.  


Even though, in essence, the valley excitation switches this does not imply intervalley scattering because the $K$ and $K'$'s excited are separated by a lattice vector and are hence not the forward and backwards propagating modes that lie within the same BZ. In summary, the orange over blue to blue over orange modal coupling is ensured by the time-reversal relationship between the two interfaces. Therefore, there is still conservation of chirality throughout this four-region structured domain.  The time-reversal relationship between the interfaces is crucial in allowing the third lead to be triggered. This provides further evidence for why only two-way splitting has thus far been obtained for TRS-breaking topological systems \cite{hammer_dynamics_2013, hammer_solitonic_2013, wang_topologically_2017, wang_topological_2018}.

Comparing our design with that of a similar hexagonal network, see \cite{makwana_designing_2018, cha_experimental_2018, he_acoustic_2016, he_silicon--insulator_2018, he_two-dimensional_2019, khanikaev_two-dimensional_2017, nanthakumar_inverse_2019, ozawa_topological_2019, qiao_current_2014, schomerus_helical_2010, shen_valley-projected_2019, xia_topologically_2019, yan_-chip_2018, ye_observation_2017, cheng_robust_2016, wu_direct_2017, xia_topological_2017, zhang_manipulation_2018, qiao_electronic_2011},  we note that the chirality and/or phase velocity mismatch results in energy being redirected solely along the two vertical partitions. Additionally there is no such relationship between the blue over orange and orange over blue zigzag interfaces, see Fig. \ref{fig:interfaces}(b), (d). This conservation of chirality and phase velocity, as well as the two distinct interfaces, restricts the hexagonal structures to two-way energy-splitting \cite{cha_experimental_2018, he_acoustic_2016, he_silicon--insulator_2018, he_two-dimensional_2019, khanikaev_two-dimensional_2017, nanthakumar_inverse_2019, ozawa_topological_2019, qiao_current_2014, schomerus_helical_2010, shen_valley-projected_2019, xia_topologically_2019, yan_-chip_2018, ye_observation_2017, cheng_robust_2016, wu_direct_2017, xia_topological_2017, zhang_manipulation_2018, qiao_electronic_2011}. 

A comprehensive pictorial comparison between the $C_{2v}$, $C_{4v}$ cases described herein and the, more common, topologically nontrivial and trivial hexagonal examples described in \cite{makwana_designing_2018} is shown in the following page.

 \clearpage
 \newpage
 

\begin{figure} [h!]
		\includegraphics[width=1.00\textwidth]{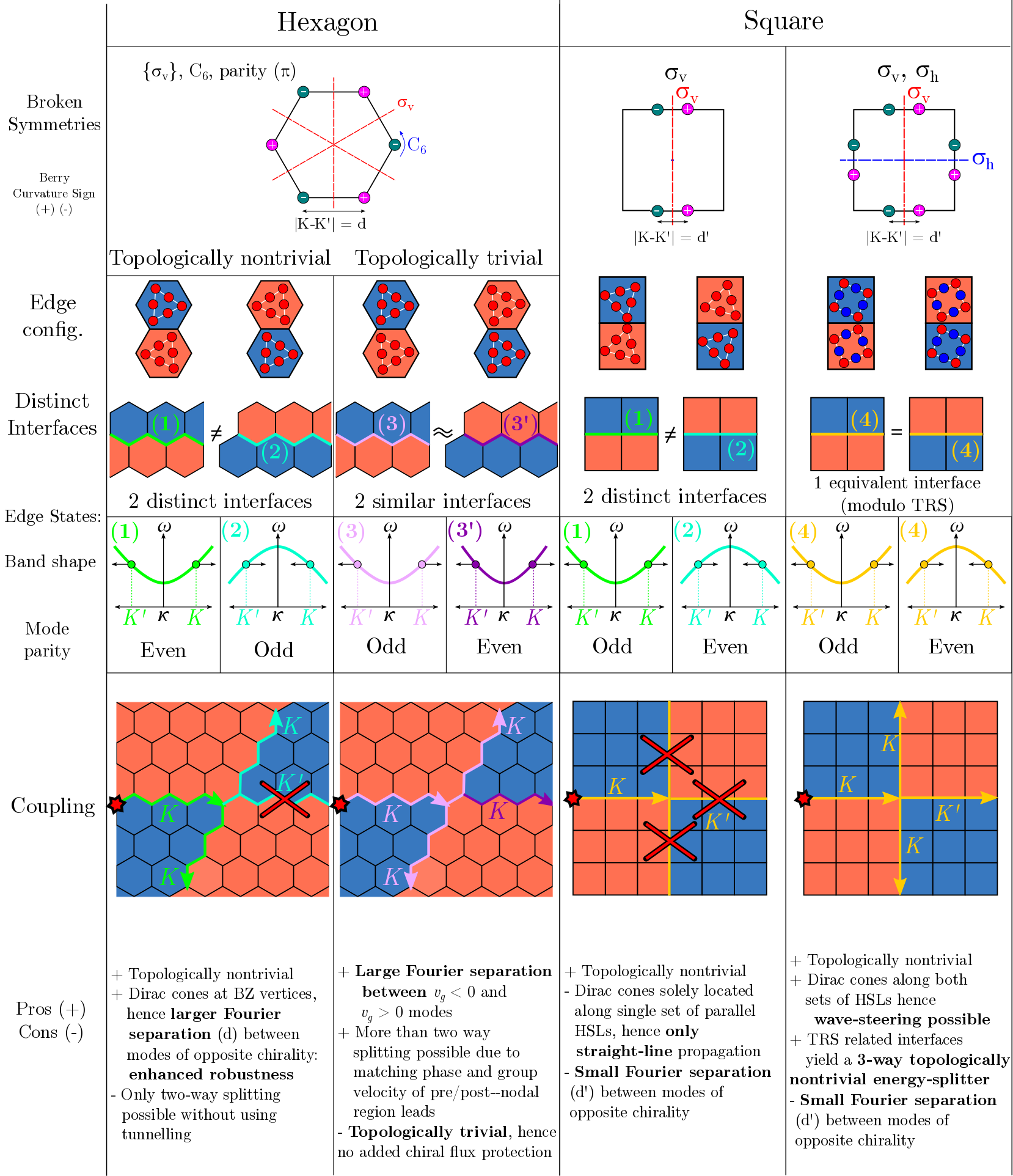}
\label{fig:summary_table}
\end{figure}

\clearpage
\section{Concluding remarks}
\label{sec:conclusion}
We have demonstrated how to geometrically engineer the first-ever broadband three-way energy-splitter. This novel paradigm adds a degree of freedom unavailable to all current designs; namely, the hexagonal valley-Hall energy-splitters \cite{cha_experimental_2018, he_acoustic_2016, he_silicon--insulator_2018, he_two-dimensional_2019, khanikaev_two-dimensional_2017, nanthakumar_inverse_2019, ozawa_topological_2019, qiao_current_2014, schomerus_helical_2010, shen_valley-projected_2019, xia_topologically_2019, yan_-chip_2018, ye_observation_2017, cheng_robust_2016, wu_direct_2017, xia_topological_2017, zhang_manipulation_2018, qiao_electronic_2011} and the two-way cavity guide beam-splitters \cite{zhao_photonic_2007, pustai_analysis_2004, tang_one-way_2018, prather_self-collimation_2007, shi_dispersion-based_2004, luan_periodic_2007, liu_multiple_2018, fan_waveguide_2001, bostan_design_2005-1, boscolo_y_2002-1, bayindir_photonic-crystal-based_2000}. This design is reliant upon the time-reversal relationship between the interfaces and hence serves as a paradigm for all scalar wave systems: plasmonics, photonics, acoustics, as well as, for vectorial systems such as plane-strain elasticity, surface acoustic waves and Maxwell equation systems. The additional degree of freedom afforded by this three-way energy-splitter, along with latest advancements in topological physics, will inevitably lead to a myriad of highly tunable, broadband and efficient crystalline networks. 

\emph{Acknowledgments} \textemdash Both authors would like to thank the EPSRC as well as Richard. V. Craster.


\section{Appendices}

\appendix

\section*{Characterising energy-carrying envelope and relevance to robustness}
\label{sec:C2v_envelope}
The efficacy of transmitting energy around a bend, coupling modes between different leads within a network or even transmission through a straight ZLM is contingent upon the displacement of the mode at the turning, nodal or end point. Knowledge of the long-scale envelope is especially useful for these finite length interfaces as it can used to minimise the backscattering as one has, in effect, a Fabry-P\'erot resonator. Examples of the characterisation of the energy-carrying envelope, using high-frequency homogenisation (HFH) 
[31], for the $C_{2v}$ and $C_{4v}$ cases are shown. In addition to this, the interfacial dispersion curves for a variation of the $C_{4v}$ case are derived using a Fourier-Hermite spectral method \cite{chaplain_rayleigh-bloch_2018}.

\begin{figure}
	\begin{tabular}{cc}
		(a) & (b) \\
		\includegraphics[height=0.43\linewidth]{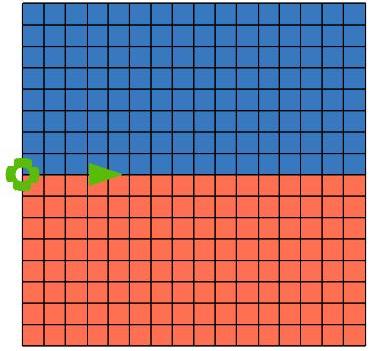} & 
		\includegraphics[width=0.425\linewidth]{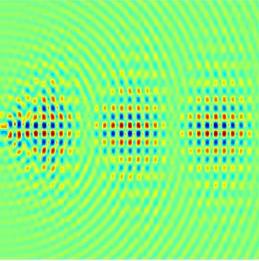} \\
		(c) & (d) \\
		\includegraphics[width=0.425\linewidth]{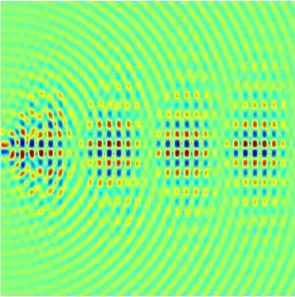} &
		\includegraphics[width=0.425\linewidth]{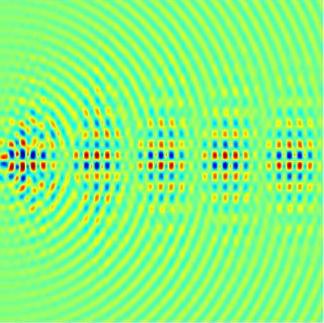}
	\end{tabular}
	\caption{Dipole source for odd ZLM excitation for $C_{2v}$ case \textemdash Long-scale modulation clearly evident from modal patterns.}
	\label{fig:C3v_ZLM_beats}
\end{figure}

To fully characterise the long-scale periodic behaviour of topological edge states along a crystal interface (Fig. \ref{fig:C3v_ZLM_beats}) we utilise HFH, applying the methodology directly in reciprocal space \cite{chaplain_rayleigh-bloch_2018}, to further bolster the plane wave expansion (PWE) method that was used to obtain the dispersion curves. This technique is a multiple scale asymptotic method, that (for non-degenerate curves with locally quadratic curvature) results in the following homogenised PDE,
\begin{align}
T_{ij}f_{0,X_{i}X_{j}} - \omega_{2}^2f_{0} = 0,
\end{align}
where $f_{0}$ is the long-scale envelope defined on the coordinate system $(X_{i},X_{j})$;  whilst the $T_{ij}$ coefficients fully encapsulates the short-scale behaviour (similar analysis can be carried over to any scalar and vectorial systems \cite{antonakakis_high_2012, antonakakis_asymptotics_2013, antonakakis_homogenization_2014, antonakakis_homogenization_2014, craster_high_2011, craster_high_2010, chaplain_rayleigh-bloch_2018}). The tensor coefficients $T_{ij}$ are geometrically dependant and, from the simple solution of the homogenised PDE, determine the envelope wavelength for a given frequency. These coefficients are determined entirely from integrated quantities of the wave-field in physical space. To avoid the need for regularisation (higher order corrections) we work in reciprocal space and calculate the $T_{ij}$'s directly, using the PWE method. Our eigenvalue problem is recast into matrix form,
\begin{equation}
\left[\matA(\boldsymbol{\kappa}) - \omega^2\matB(\boldsymbol{\kappa})\right]\mathbf{W} = 0,
\end{equation}
with the matrices $\matA, \matB$ encoding the geometry and forcing of the mass loading.

\begin{figure}[h!]
	\includegraphics[scale=0.2]{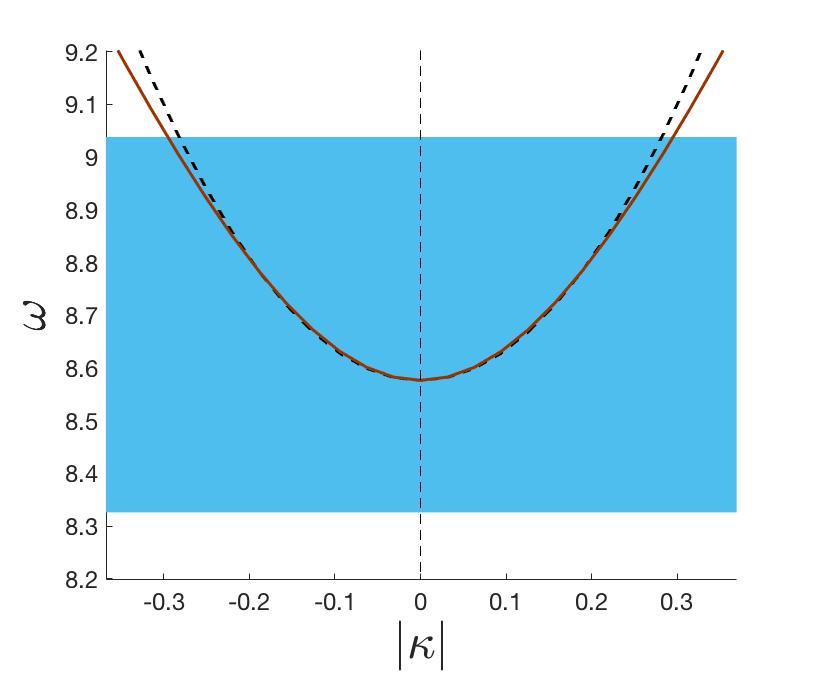}
	\caption{HFH applied to $C_{2v}$ interfacial mode \textemdash HFH asymptotics applied to the odd-parity curve (also shown in Fig. \ref{fig:C3v_edge_modes}); asymptotic curve (dashed line) computed using the  $T_{ij}$ coefficients, Eq. \eqref{eq:Tij}. An example of the characterisation of the energy-carrying envelope is shown in Fig. \ref{fig:C2v_ZLM}(b)}
	\label{fig:HFH_triangle}
\end{figure}

Expanding in the vicinity of a high symmetry point leads to the following ansatz;
\begin{align}
\begin{split}
\omega^2 &= \omega_{0}^2 + \epsilon\omega_{1}^2 + \epsilon\omega_{2}^2 + \hdots \\   
\mathbf{W} &= \mathbf{W}_{0} + \epsilon\mathbf{W}_{1} + \epsilon^2\mathbf{W}_{2} + \hdots \\
\matA &= \matA_{0} + \epsilon\sum\limits_{i}\kappa_{i}\underbrace{\frac{\partial\matA}{\partial\kappa_{i}}\bigg\vert_{\Gamma}}_{\matA^{(1)}_{i}} + \epsilon^2\sum\limits_{i,j}\kappa_{i}\underbrace{\frac{\partial^2\matA}{\partial\kappa_{i}\kappa_{j}}\bigg\vert_{\Gamma}}_{\matA^{(2)}_{ij}}\kappa_{j} + \mathcal{O}(\epsilon^{3}), 
\end{split}
\end{align}
with a similar expansion for $\matB(\boldsymbol{\kappa})$. Applying suitable solvability conditions and imposing Bloch conditions on the microscale results in the following 
tensor coefficients $T_{ij}$, 
\begin{align}
\begin{split}
(\mathbf{\tilde{W}}_{1})_{i} \equiv \frac{(\mathbf{W}_{1})_{i}}{\kappa_{i}} = -\left[\matA_{0}-\omega_{0}^2\matB_{0}\right]^{+}\left(\matA^{(1)}_{i} - \omega_{0}^2\matB^{(1)}_{i}\right)\mathbf{W}_{0} \\ \\
T_{ij} = \frac{\mathbf{W}_{0}^{\dagger}(\matA^{(1)}_{i}-\omega_{0}^2\matB^{(1)}_{i})(\mathbf{\tilde{W}}_{1})_{i}+\frac{1}{2}\mathbf{W}_{0}^{\dagger}(\matA^{(2)}_{ij} - \omega_{0}^2\matB^{(2)}_{ij})\mathbf{W}_{0}}{\mathbf{W}_{0}^\dagger\matB_{0}\mathbf{W}_{0}},
\end{split}
\label{eq:Tij}
\end{align}
where $\omega_{0}$ and $\mathbf{W}_{0}$ are the solutions obtained from the PWE method, and $[\hdots]^{+}$ denotes the pseudoinverse. The asymptotics for the $C_{2v}$ case are plotted in Fig~\ref{fig:HFH_triangle}; the explicit characterisation of the envelope is shown in Fig. \ref{fig:C2v_ZLM}(b). 


\bibliographystyle{apsrev4-1}

\end{document}